\documentclass[traditabstract]{aa} 
\usepackage{graphicx}
\usepackage{amssymb}
\usepackage[varg]{txfonts}

\begin{document}

   \title{Star formation laws in Luminous Infrared Galaxies}

   \subtitle{New observational constraints on models\thanks{Based on observations carried out with the IRAM 30m telescope. 
IRAM is supported by INSU/CNRS (France), MPG (Germany) and IGN (Spain)}}

   \author{S.~Garc\'{\i}a-Burillo\inst{1}
			\and
	   A.~Usero\inst{1}		   
			\and			
	   A.~Alonso-Herrero\inst{2,3}
	             \and		
	   J.~Graci\'a-Carpio\inst{4}
			\and
	   M.~Pereira-Santaella\inst{2}
	             \and
         L.~Colina\inst{2} 
	   	      \and			
	  P.~Planesas\inst{1, 5}
	   		 \and 		
	   S.~Arribas\inst{2}}

   \institute{Observatorio Astron\'omico Nacional (OAN)-Observatorio de Madrid, Alfonso XII, 3, 28014-Madrid, Spain \\
			  \email{s.gburillo@oan.es}
	 \and
	 Centro de Astrobiolog\'{\i}a (CSIC-INTA), Ctra de Torrej\'on a Ajalvir, km 4, 28850 Torrej\'on de Ardoz, Madrid, Spain
	\and
	 Instituto de F\'{\i}sica de Cantabria, CSIC-UC, Avenida de los Castros s/n, 39005 Santander, Spain
	\and
	 Max-Planck-Institut f\"ur extraterrestrische Physik, Giessenbachstrasse 1, 85748 Garching, Germany	 
	\and
         ESO \& Joint ALMA Observatory, Alonso de Cordova 3107, Vitacura, Santiago, Chile}
	\date{Received ---; accepted ----}

\abstract{
   The observational study of star formation relations in galaxies is central to unraveling the related physical processes that are at work on both local and global scales. It is still debated whether star formation can be described by a universal law that remains valid in different populations of galaxies.
   We wish to expand the sample of extreme starbursts, represented by local luminous and ultra-luminous infrared galaxies (LIRGs 
   and  ULIRGs), with high quality observations in the 1--0 line of HCN, which is taken as a proxy for the dense molecular gas content. The new data presented in 
   this work allow us to enlarge in particular the number of LIRGs studied in HCN by a factor 3 compared to previous works. The chosen LIRG sample has a range of HCN luminosities that partly overlaps with that of the normal galaxy population. We study if a universal law         
   can account for the star formation relations observed for the dense molecular gas in normal star forming galaxies and extreme 
   starbursts and explore the validity of different  theoretical prescriptions of the star formation law. 
   We have used the IRAM 30m telescope to observe a sample of 19 LIRGs in the 1--0 lines of CO, HCN and HCO$^+$. The galaxies have been extracted from a sample of local LIRGs with available high-quality and high-resolution images obtained at optical, near and mid IR wavelengths, which probe the star formation activity. We have thus derived the star formation rates  using different tracers and 
  determined the sizes of the star forming regions in all the targets. 
  
   The analysis of the new data proves that the efficiency of star formation in the dense molecular gas (SFE$_{\rm dense}$) of extreme starbursts is a factor 3--4 higher compared to 
   normal galaxies. Kennicutt-Schmidt  (KS) power laws have also been derived. We find a duality in KS laws that is further reinforced if we account for the 
   likely different conversion factor for HCN ($\alpha^{\rm HCN}$) in extreme starbursts and for the unobscured star formation rate in normal galaxies.  This result extends  to the higher molecular densities probed by HCN lines the more extreme bimodal behavior of star 
 formation laws, derived from CO molecular lines by two recent surveys.
  
  We have confronted our observations with the predictions of theoretical 
   models in which the efficiency of star formation is determined by the ratio of a \emph{constant} star formation rate per free-fall time (SFR$_{\rm ff}$) to the 
   local free-fall time (t$_{\rm ff}$). We find that it is possible to fit the observed differences in the SFE$_{\rm dense}$ between normal galaxies and LIRGs/ULIRGs using a common constant SFR$_{\rm ff}$ and a set of physically acceptable HCN densities, but only if SFR$_{\rm ff}$$\sim$0.005--0.01 and/or if $\alpha^{\rm HCN}$ is  a factor of $\sim$a few lower than our favored values. Star formation recipes that explicitly depend on the galaxy global dynamical time scales do not significantly improve the fit to the new HCN data presented in this work.}
   \maketitle
%

\section{Introduction}\label{Introduction}

The quest for a quantitative description of how the star formation rate should depend on a set of galaxy parameters (gas and stellar content, gas 
kinematics, environment) is a fundamental though not fully understood problem. Observational constraints are central to understanding 
the role that small-scale ({\it local}) and large-scale ({\it global}) galaxy properties play in star formation relations. This is a requisite to 
validating the recipes of star formation used in numerical simulations of galaxy evolution.

Based on simple  theoretical grounds Schmidt~(\cite{Sch59}) proposed that the star formation rate (SFR) per unit volume ($\rho_{\rm SFR}$) 
should be a power law of index $N$ of the gas volume density ($\rho_{\rm gas}$): $\rho_{\rm SFR} \propto \rho_{\rm gas}^{N}$. 
Schmidt~(\cite{Sch59}) predicted that $N$ should be around 2. If we suppose that the relevant time scale for star formation is the 
{\it local} free-fall time (t$_{\rm ff}$), in this case 
$\rho_{\rm SFR} \propto \rho_{\rm gas}$/$t_{\rm ff}\propto \rho_{\rm gas}^{1.5}$. On the condition that the gas scale-height is constant among 
galaxies, a non-trivial additional assumption, the equivalent of the Schmidt law in terms of the corresponding surface densities of SFR 
($\Sigma_{\rm SFR}$) and gas ($\Sigma_{\rm gas}$) is $\Sigma_{\rm SFR} \propto \Sigma_{\rm gas}^{1.5}$. 

Krumholz \&McKee~(\cite{Kru05}) (see also Krumholz, McKee \& Tumlinson~\cite{Kru09}) developed a model 
that successfully describes how star formation takes  place in a highly turbulent molecular medium. Their scenario pictures 
star formation as a very inefficient process that takes place in small subregions of supersonic turbulent and virialized 
molecular clouds. The surface density of the star formation rate is in any case expected to scale as a power law of the mean density of the gas with 
an index close to 1.5. Star formation in Krumholz \&McKee's formulation is regulated by processes mostly within the clouds. 
The relevant time scale for these models, the free-fall time for cloud collapse, is thus a {\it local} time scale. 
In this type of  formulation, star formation laws should not depend explicitly on global dynamical time scales or be 
determined by galactic-scale phenomena. Instead, large-scale processes are thought to be responsible for  making the gas 
assembly {\it prior} to star formation in galaxies. Alternatively, other forms of the star formation laws adduce the 
importance of including explicitly {\it global} dynamical time scales  t$_{\rm dyn}$ (Elmegreen~\cite{Elm97};  Silk~\cite{Sil97}; Kennicutt~\cite{Ken98}; 
Elmegreen~\cite{Elm02}; Boissier et al.~\cite{Boi03}; Tasker \& Tan~\cite{Tas09}; Tan~\cite{Tan10}; Genzel et al.~\cite{Gen10}). 
The star formation rate density $\Sigma_{\rm SFR}$ should then scale as $\propto$$\Sigma_{\rm gas}$/t$_{\rm dyn}$. 
It is yet unclear if the two forms of the star formation laws are fully equivalent both in normal galaxies and in mergers. 
Kennicutt~(\cite{Ken98}) found that if t$_{\rm dyn}$ is the {\it average} orbital dynamical time scale, both formulations are 
equally successful at describing star formation laws in the disks of normal galaxies. The physical framework for this equivalence 
relies on the fact that normal galaxy disks, viewed as dynamically relaxed stable systems, fulfill the Toomre stability criterium 
(Toomre~\cite{Too64}). It is nevertheless questionable if this equivalence should also hold in 
extreme starbursts which are often hosted by merging systems.

During the last decades observers have been using different proxies for  $\Sigma_{\rm SFR}$ and $\Sigma_{\rm gas}$ to 
validate the predictions of star formation law models. These observations have been conducted on different galaxy samples, 
including normal galaxies and extreme starbursts at different redshifts. They also encompass a 
wide range of spatial resolutions, which allow not only global measurements but also studies of spatially resolved objects.  
The different observers do not find a single exponent fitting the entire star formation relation. Kennicutt~(\cite{Ken98}) finds an index $n$=1.40$\pm$0.15 using 
CO and HI as gas tracers and a $\sim$100 galaxy sample that includes normal galaxies, nuclear starbursts and a few LIRGs ($\sim$5) at z=0. Yao et al.~(\cite{Yao03}) find 
an index $n$=1.40$\pm$0.30 in their CO survey of IR luminous galaxies.
Bouch\'e et al.~(\cite{Bou07}), by including submillimeter galaxies (SMGs) and using a lower CO conversion factor in z=0 ULIRGs, determines a higher index for the Kennicutt-Schmidt (KS) law $n$$\sim$1.7. The high-resolution survey of 18 normal galaxies
published by Bigiel et al.~(\cite{Big08}) and Leroy et al.~(\cite{Ler08}) found a linear KS relation above a gas surface density threshold of 10$M_{\sun}$pc$^{-2}$, which determines the transition from atomic to molecular gas, here traced by CO lines.

\begin{table*}[!ht]
\centering
\caption{\label{t1}Main properties of the new sample of LIRGs.}
 \begin{tabular}{lccccccc} 
\hline\hline 
\noalign{\smallskip} 
         galaxy &      $RA_{J2000}$ &     $DEC_{J2000}$ &       d &     $z_{CO}$ &    $L_\mathrm{FIR}$ &     $L_\mathrm{IR}$ &   $d_\mathrm{SF}$\tablefootmark{a} \\
                & $hh$:$mm$:$ss.ss$ &  $dd$:$mm$:$ss.s$ &     Mpc &         &    $10^{11}L_\odot$ &    $10^{11}L_\odot$ &            arcsec \\
\noalign{\smallskip} 
\hline 
\noalign{\smallskip}

         NGC~23 &       00:09:53.40 &       +25:55:25.9 &    60.1 & 0.01521 &                0.79 &                1.14 &               7.1 \\
  MCG+12-02-001 &       00:54:04.00 &       +73:05:05.5 &    65.8 & 0.01587 &                2.04 &                2.88 &               4.7 \\
      III~Zw~35 &       01:44:30.54 &       +17:06:08.8 &   114.5 & 0.02753 &                3.79 &                4.16 &              10.0 \\
       UGC~1845 &       02:24:07.98 &       +47:58:11.0 &    64.1 & 0.01566 &                0.95 &                1.25 &               4.3 \\
       NGC~1614 &       04:34:00.02 &       -08:34:44.9 &    66.6 & 0.01575 &                3.04 &                4.50 &               3.1 \\      
       UGC~3351 &       05:45:47.89 &       +58:42:03.9 &    63.1 & 0.01480 &                1.52 &                1.78 &               4.5 \\
       NGC~2388 &       07:28:53.44 &       +33:49:08.7 &    60.5 & 0.01369 &                1.38 &                1.86 &               5.2 \\
  MCG+02-20-003 &       07:35:43.43 &       +11:42:35.0 &    73.7 & 0.01655 &                1.13 &                1.43 &               8.1 \\
       NGC~2623 &       08:38:24.08 &       +25:45:16.6 &    82.7 & 0.01848 &                3.45 &                3.96 &               4.0 \\
       NGC~3110 &       10:04:02.05 &       -06:28:29.4 &    77.8 & 0.01697 &                1.82 &                2.29 &               6.2 \\
IRAS~10173+0828 &       10:20:00.18 &       +08:13:33.8 &   220.1 & 0.04900 &                6.15 &                7.75 &              10.0 \\
         IC~860 &       13:15:03.52 &       +24:37:07.9 &    58.9 & 0.01289 &                1.37 &                1.47 &               2.9 \\
       NGC~5653 &       14:30:10.48 &       +31:12:55.7 &    53.4 & 0.01191 &                0.84 &                1.09 &               7.6 \\
       NGC~5936 &       15:30:00.86 &       +12:59:21.1 &    59.0 & 0.01334 &                0.80 &                1.10 &               3.4 \\
IRAS~17138-1017 &       17:16:35.79 &       -10:20:39.4 &    74.2 & 0.01734 &                1.78 &                2.52 &               4.6 \\
IRAS~18090+0130 &       18:11:38.39 &       +01:31:40.0 &   123.5 & 0.02891 &                3.26 &                4.01 &              10.0 \\
       NGC~6701 &       18:43:12.46 &       +60:39:12.0 &    54.5 & 0.01315 &                0.79 &                1.04 &               4.6 \\
       II~Zw~96 &       20:57:23.84 &       +17:07:39.8 &   161.0 & 0.03629 &                7.40 &                9.30 &              10.0 \\
       NGC~7591 &       23:18:16.28 &       +06:35:08.9 &    65.4 & 0.01654 &                0.87 &                1.12 &               2.9 \\

\noalign{\smallskip} 
\hline 
\end{tabular}
\tablefoot{
\tablefootmark{a}{$d_\mathrm{SF}$ is the source size (major axis) of the star forming region in arscec determined from high-resolution optical/NIR images of hydrogen recombination  lines (H$\alpha$, Pa$\alpha$).} 
}
 \end{table*}

More recently the CO surveys of Daddi et al.~(\cite{Dad10}) and Genzel et al.~(\cite{Gen10}) presented evidence that normal galaxies and mergers (LIRGs, ULIRGs and SMGs) occupy different regions in the molecular gas mass versus star formation rate plane. These results suggest the existence of a bimodality in star formation laws, where normal galaxies show a 4--10 longer depletion time scales compared to mergers. While the KS relation remains mostly linear inside each galaxy population, bimodality introduces a discontinuity in the two-function power law. These authors discussed that when global dynamical time scales are included a universal star formation law is obtained. Based on a study of far-infrared fine structure lines,  Graci\'a-Carpio et al.~(\cite{Gra11}) concluded that extreme starbursts display significantly enhanced line deficits. The results of Graci\'a-Carpio et al.~(\cite{Gra11}) provide additional support for the existence of different properties of the interstellar medium in normal star forming galaxies and extreme starburst systems.

Krumholz  \& Thompson~(\cite{Kru07b}) (see also Narayanan et al.~\cite{Nar08})  studied how the power index of KS laws determined from observations should change 
depending on critical density of the tracer used to probe the star forming gas, a prediction confirmed by observations done in several dense gas tracers (Narayanan et al.~\cite{Nar05}; 
 Bussmann et al.~\cite{Bus08}; Bayet et al.~\cite{Bay09}; Juneau et al.~\cite{Jun09}). In the particular case of HCN(1--0), only in 
galaxies where the average gas density exceeds a few $10^{4}\,\rm cm^{-3}$ (i.e., the effective critical density of the 
HCN J=1--0 line), we would start recovering the expected superlinear behavior of the \emph{universal} KS law derived by 
Krumholz \& McKee~(\cite{Kru05}).

Observations of HCN lines have been used to study dense molecular gas properties in samples of normal spirals, 
LIRGs, ULIRGs, QSOs and molecular gas-rich high-$z$ galaxies (e.g., Gao \&\ Solomon~\cite{Gao04a, Gao04b}; Solomon \&\ 
Vanden Bout~\cite{Sol05}, Wagg et al.~\cite{Wag05}, Riechers et al.~\cite{Rie07}, Graci\'a-Carpio et al.~\cite{Gra06,Gra08}). 
By contrast to CO line observations, which are sensitive to the global molecular gas content, HCN observations are a much better tracer of the dense molecular phase, which is more directly related to star formation, and can thus put stringent constraints on star formation models. The
most important result presented by Gao \&\ Solomon in their seminal papers is the discovery of a tight correlation between the
infrared and the HCN(1--0) luminosities over three orders of magnitude in $L_{\rm{IR}}$. This result was interpreted as evidence 
that star formation is the main power source in ULIRGs. The linearity of the tight correlation between the IR and the HCN(1--0) 
luminosities implies that the star formation efficiency measured with respect to the dense molecular gas content 
(SFE$_{\rm{dense}} \propto L_{\rm{IR}}/L'_{\rm{HCN(1-0)}}$) is constant in all galaxies, independently of $L_{\rm{IR}}$.

 More recent results have partly questioned this picture, however. Graci\'a-Carpio et al.~(\cite{Gra08}) published observations
 made with the IRAM 30m telescope of the J=1--0 and 3--2 lines of HCN and HCO$^{+}$ used to probe the dense molecular gas 
 content of a sample of 17 LIRGs and ULIRGs. These observations were also used to derive a new version of the power law 
 describing the correlation between $L_{\rm FIR}$ and $L'_{\rm HCN(1-0)}$ from normal galaxies ($L_{\rm IR} < 10^{11}\,L_
 {\sun}$) to high-$z$ galaxies. The results of Graci\'a-Carpio et al.~(\cite{Gra08}) indicated that the $L_{\rm FIR}/L'_{\rm HCN(1-0)}$ ratio, 
 taken as proxy for SFE$_{\rm dense}$, is a factor $\sim$2--3 higher in IR luminous targets compared 
 to normal galaxies. Furthermore, Graci\'a-Carpio et al.~(\cite{Gra08}) 
 found, based on a multiline analysis of HCN and HCO$^{+}$ data, that $X_{\rm HCN}$ is probably about 3 times lower at high $L_{\rm FIR}$. 
 Taken together, these findings suggest that SFE$_{\rm dense}$ may be up to an order of magnitude higher in 
 extreme starbursts (LIRGs, ULIRGs) than in normal galaxies.

 Compared to normal galaxies and 
the most extreme starbursts represented by ULIRGs,  LIRGs can constitute the 
transition point in the star formation laws of galaxies. The main purpose of this paper is to enlarge the sample of LIRGs with high quality HCN(1--0) observations. 
This is a key to testing if the {\it observed} bimodality of star formation laws in galaxies 
derived from CO, can be extended to the  higher density regime probed by HCN. We use these results to explore the validity of different  star 
formation law theoretical prescriptions which are currently debated. To this aim we have used the 30m telescope to observe a sample of 19 LIRGs 
in the 1--0 lines of CO, HCN and HCO$^+$. Most of the targets have been extracted from the complete sample of local (d$<$78~Mpc) 
LIRGs studied by Alonso-Herrero et al.~(\cite{Alo06}). All the targets in this sample have high-quality and high-resolution 
imaging at optical, NIR and MIR wavelengths (obtained with HST, VLT, CAHA and Spitzer). This allows us to derive 
 accurately the sizes of the star forming regions and the star formation rates for all the targets. The new data presented 
in this work allow us to expand the number of LIRGs studied in HCN-HCO$^+$ lines by more than a factor 3 compared to previous works 
(9 LIRGs in Graci\'a-Carpio et al.'s sample).

\begin{figure*}[!ht]
\centering
\scalebox{0.63}{\includegraphics{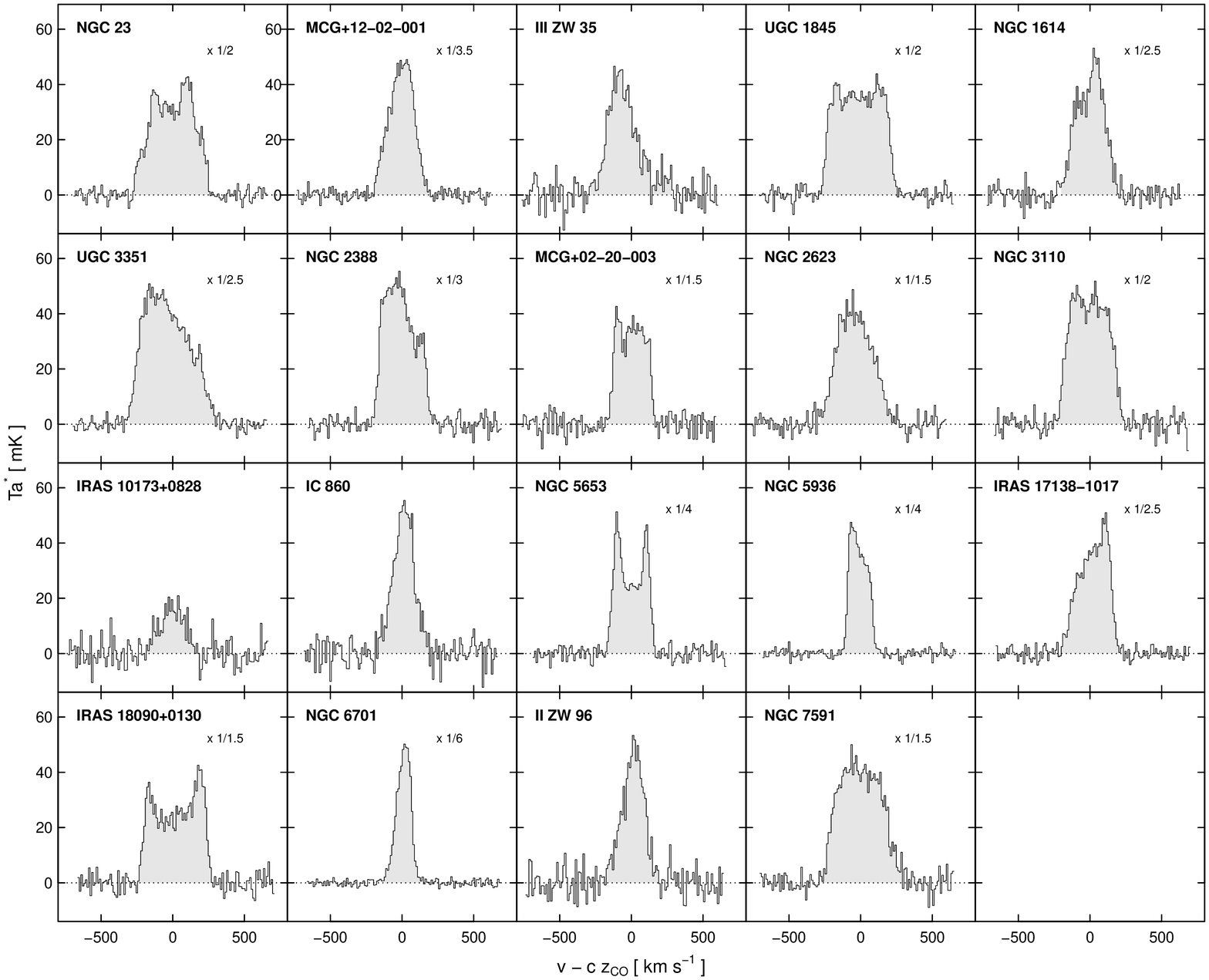}} 
\caption{{\bf a)} CO(1--0) spectra observed with the IRAM 30m telescope in a sample of 19 LIRGs. Spectra appear from top to bottom and from left to right in order of
increasing right ascension of the source.
To ease the display some of the line intensities have been scaled by the factor indicated in the panel. Grey-filled
histograms highlight the velocity range used to calculate the baseline fitting and the line areas.}
\label{spectra}
\end{figure*}

\begin{figure*}[!ht]
\centering
\scalebox{0.63}{\includegraphics{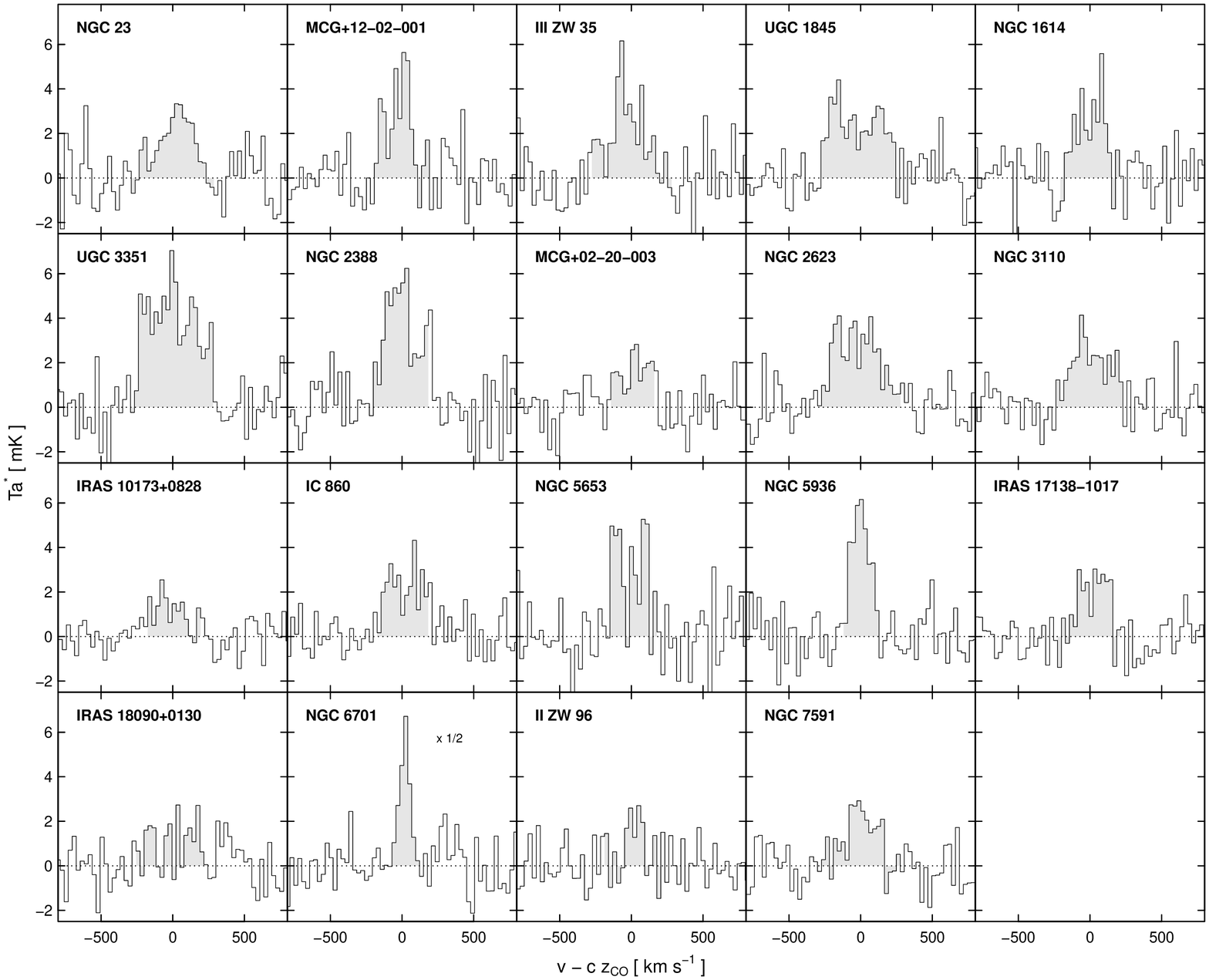}} 
\addtocounter{figure}{-1}
\caption{{\bf b)} Same as {\bf a)} but showing the HCN(1--0) spectra.}
\end{figure*}

\begin{figure*}[!ht]
\centering
\scalebox{0.63}{\includegraphics{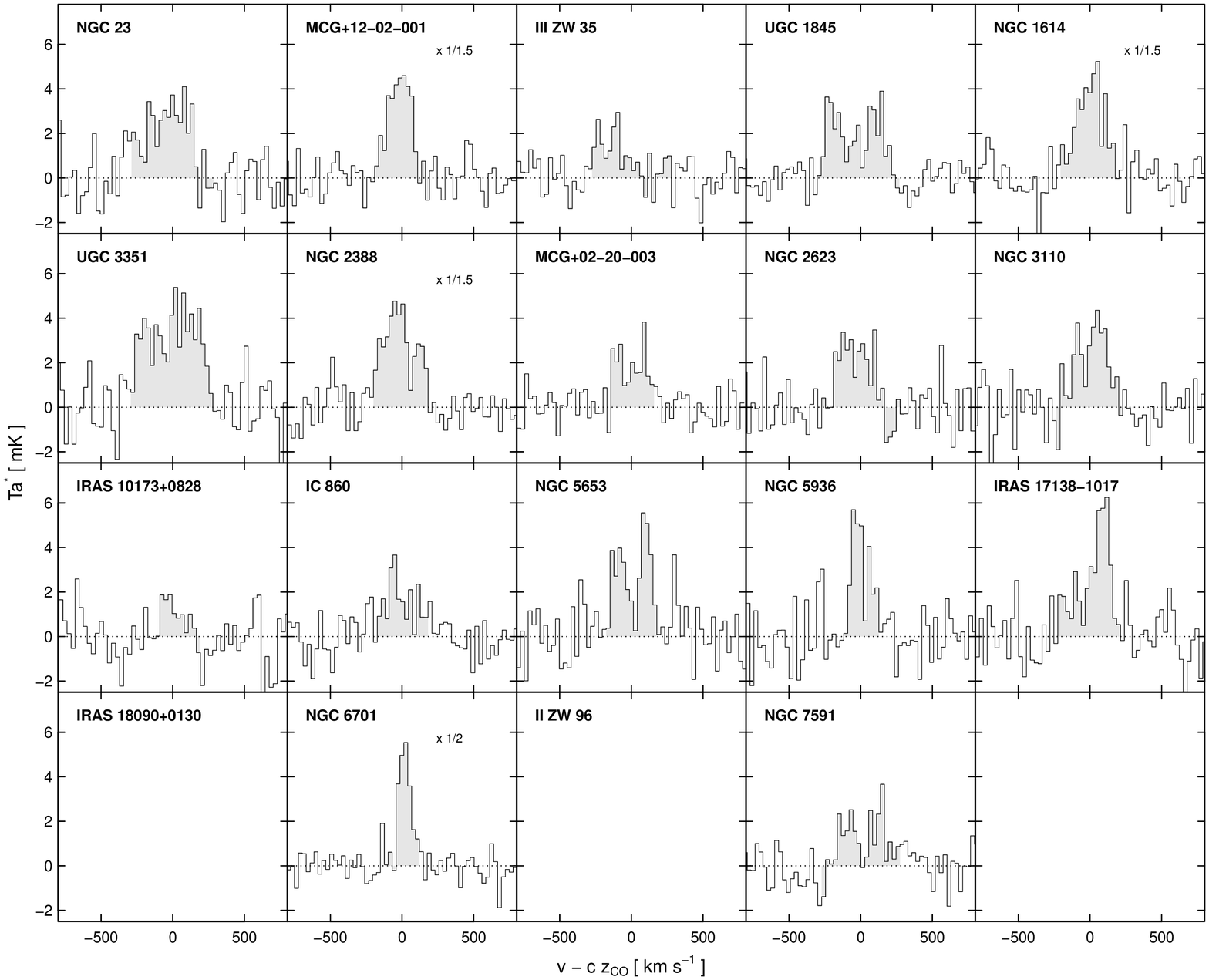}} 
\addtocounter{figure}{-1}
\caption{{\bf c)} Same as {\bf a)} but showing the HCO$^{+}$(1--0) spectra.}
\end{figure*}


\section{The sample of LIRGs}\label{Sample}

We have obtained new observations for a sample of 19 LIRGs. Out of these, 
fourteen have been extracted from the sample of
nearby LIRGs studied by Alonso-Herrero et al.~(\cite{Alo06}).  
This is a volume-limited sample (v$<$5200~km/s, distances of
between 35 and 78~Mpc) drawn from the IRAS Revised Bright  
Galaxy Sample (RBGS) analyzed by Sanders et al.~(\cite{San03}). 
Alonso-Herrero et al.~(\cite{Alo06}) required the logarithm of 
the total IR luminosity to be $\geq$11.05~L$_{\odot}$ and the galaxies
to be at Galactic 
latitude b$>$10$^{\circ}$. The remaining five LIRGs were selected
to populate the high end of the IR luminosity distribution of LIRGs.
 For the Alonso-Herrero et al. LIRGs we have obtained HST/NICMOS Pa$\alpha$ imaging (Alonso-Herrero
et al.~\cite{Alo01, Alo02, Alo06}),   
as well as ground-based optical integral field spectroscopy using CAHA/PMAS and
VLT/VIMOS covering the H$\alpha$ spectral region 
(Alonso-Herrero et al.~\cite{Alo09}; Rodr\'{\i}guez Zaur\'{\i}n
  et al.~\cite{Rod11}). In addition there are mid infrared (MIR) Spitzer
imaging observations for the entire sample from the GOALS sample
  (Armus et al.~\cite{Arm09}).  This means that the star
formation rates (SFR) of the sample are very well characterized from
hydrogen recombination lines (H$\alpha$, Pa$\alpha$) and MIR
(24$\mu$m) luminosities on the same angular scales probed by the IRAM
observations. We also have a good estimate  
of the typical size of the star forming regions in all the galaxies of
our sample, thanks to the  
available high-resolution imaging (H$\alpha$, Pa$\alpha$). This
minimizes the errors when both $\Sigma_{\rm SFR}$ and  
$\Sigma_{\rm dense}$, required to derive KS laws, are evaluated. The
estimated sizes are typically  
$<$10$\arcsec$ in all sources, equivalent to 1.7--3.6~kpc for the given range of distances. This guarantees that the total emission of the molecular gas is 
contained  in the 30m beam ($\sim$28$\arcsec$ at the frequency of the
HCN(1--0) line).   

Additionally, in this paper we use a compiled sample of 108
galaxies with published FIR and HCN(1--0) observations in the
literature.  
When available, we also include HCO$^+$(1--0) data. Besides the new
data obtained in LIRGs for this work we include  
the data of normal galaxies, LIRGs and ULIRGs of Gao \&
Solomon~(\cite{Gao04a}), data from the sample of  
infrared-excess Palomar-Green QSOs (Evans et al.~\cite{Eva06}) and
data from high-$z$ galaxies with available HCN  
observations (see Gao et al~\cite{Gao07} and references therein).
Following the same approach adopted by  
Graci\'a-Carpio et al.~(\cite{Gra08}), in those high-$z$ sources where
HCN(1--0) are not available, $L'_{\rm HCN(1-0)}$  
was derived assuming that their rotational line luminosity ratios are
$L'_{\rm HCN(2-1)}/L'_{\rm HCN(1-0)} = 0.7$, similar 
to the mean value measured by Krips et al.~(\cite{Kri08}). For the $z
\simeq 4$ quasar APM 08279+5255 we adopt 
 $L'_{\rm HCN(5-4)}/L'_{\rm HCN(1-0)} = 0.3$, assuming the physical
 conditions derived in Garc\'{\i}a-Burillo et al.~(\cite{Gar06}).

 
\begin{table*}[hbt!]
\caption{\label{t2}Results derived from the IRAM 30m telescope observations.}
\centering
\begin{tabular}{lrrrrrr} 
\hline\hline 
\noalign{\smallskip} 
         galaxy &     \multicolumn{2}{c}{$I_\mathrm{CO}$\tablefootmark{a}} &    \multicolumn{2}{c}{$I_\mathrm{HCN}$\tablefootmark{a}} &  \multicolumn{2}{c}{$I_\mathrm{HCO^+}$\tablefootmark{a}} \\
                &         \multicolumn{2}{c}{K~km~s$^{-1}$} &         \multicolumn{2}{c}{K~km~s$^{-1}$} &         \multicolumn{2}{c}{K~km~s$^{-1}$} \\
\noalign{\smallskip} 
\hline 
\noalign{\smallskip} 
         NGC~23 &   28.36 &                          (0.32) &    0.81 &                          (0.13) &    1.08 &                          (0.13) \\
  MCG+12-02-001 &   36.50 &                          (0.39) &    0.89 &                          (0.10) &    1.34 &                          (0.10) \\
      III~Zw~35 &   10.43 &                          (0.35) &    1.08 &                          (0.15) &    0.50 &                          (0.10) \\
       UGC~1845 &   32.90 &                          (0.31) &    1.29 &                          (0.13) &    1.00 &                          (0.13) \\
       NGC~1614 &   27.16 &                          (0.47) &    0.79 &                          (0.11) &    1.56 &                          (0.11) \\
       UGC~3351 &   46.51 &                          (0.40) &    2.17 &                          (0.13) &    1.81 &                          (0.13) \\
       NGC~2388 &   42.60 &                          (0.52) &    1.51 &                          (0.11) &    1.69 &                          (0.11) \\
  MCG+02-20-003 &   14.08 &                          (0.30) &    0.52 &                          (0.10) &    0.54 &                          (0.10) \\
       NGC~2623 &   19.02 &                          (0.34) &    1.30 &                          (0.11) &    0.67 &                          (0.12) \\
       NGC~3110 &   31.37 &                          (0.46) &    0.92 &                          (0.10) &    0.87 &                          (0.10) \\
IRAS~10173+0828 &    3.52 &                          (0.26) &    0.43 &                          (0.11) &    0.29 &                          (0.11) \\
         IC~860 &    9.61 &                          (0.25) &    0.80 &                          (0.10) &    0.55 &                          (0.10) \\
       NGC~5653 &   36.56 &                          (0.57) &    1.00 &                          (0.10) &    1.00 &                          (0.10) \\
       NGC~5936 &   27.44 &                          (0.30) &    0.94 &                          (0.08) &    0.73 &                          (0.10) \\
IRAS~17138-1017 &   27.36 &                          (0.35) &    0.58 &                          (0.09) &    1.12 &                          (0.13) \\
IRAS~18090+0130 &   19.06 &                          (0.34) &    0.51 &                          (0.11) &    ---- &                           ----- \\
       NGC~6701 &   36.17 &                          (0.32) &    1.11 &                          (0.17) &    1.15 &                          (0.08) \\
       II~Zw~96 &    8.97 &                          (0.32) &    0.35 &                          (0.09) &    ---- &                           ----- \\      
       NGC~7591 &   24.66 &                          (0.31) &    0.61 &                          (0.13) &    0.57 &                          (0.13) \\

\noalign{\smallskip} 
\hline 
\end{tabular} 
\tablefoot{
\tablefootmark{a}{All velocity integrated intensities are in T$_{\rm a}^*$ scale. We list 1-$\sigma$ errors within brackets.} 
}
\end{table*}

\section{Observations}\label{Observations}

The new molecular line observations of the sample of 19 LIRGs described in Sect.~\ref{Sample} were carried out in one 
observing run on June 2008 with the IRAM 30m telescope at Pico de Veleta (Spain). We tuned the 3\,mm and 1\,mm SIS 
receivers of the 30m telescope to the redshifted frequencies of the HCN(1--0), HCO$^+$(1--0), CO(1--0) and CO(2--1) 
lines. The beam sizes of the 30m telescope range from  FWHM$\sim$28$\arcsec$ at  88\,GHz to  
$\sim$11$\arcsec$ at  230\,GHz). We covered a velocity range $\rm 1300-1800\,km\,s^{-1}$ for the 3\,mm lines and 
$\rm 1200\,km\,s^{-1}$ for the 1\,mm lines. We used the wobbler switching mode to make baselines flatter. 
System temperatures during the observations were typically $\sim$100--130\,K at  3\,mm and $\sim$300--500\,K at 1\,mm. Receivers were 
used in single side-band mode (SSB), with a high rejection of the image band: $>$12\,dB at 1\,mm and $>$20\,dB at 3\,mm. 
The calibration accuracy is estimated to be better than 20$\%$. We checked the pointing of the  telescope every 1.5 hours 
by observing nearby continuum sources; the average rms pointing error was  2$\arcsec$ during the observing run. 

Throughout the paper, velocity-integrated line intensities ($I$) are given in antenna temperature scale, $T_{\rm a}^{*}$. The $T_
{\rm a}^{*}$ scale relates to the main beam temperature scale, $T_{\rm mb}$, by the equation $T_{\rm mb} = (F_{\rm eff}/B_{\rm 
eff}) T_{\rm a}^{*}$, where $F_{\rm eff}$ and $B_{\rm eff}$ are the forward and beam efficiencies of the telescope at a given 
frequency. For the IRAM 30m telescope $F_{\rm eff}/B_{\rm eff} = 1.22$ (1.57) at 87-88\,GHz (230\,GHz) and $S/T_{\rm mb} = 4.95$
\,Jy\,K$^{-1}$. The velocity windows used to derive $I_{\rm HCO^{+}}$  and  $I_{\rm HCN}$ have been 
defined from the corresponding higher signal-to-noise CO line profiles. We have subtracted linear baselines to the individual scans and eliminated
those showing instabilities requiring higher order baselines. We show in Fig.~\ref{spectra} the  ensemble of HCN(1--0), HCO$^+$(1--0) and 
CO(1--0) spectra used in this work. Molecular line luminosities ($L'$) were computed in units of 
$L' = \rm K\,km\,s^{-1}\,pc^{2}$ as defined by Gao \& Solomon~(\cite{Gao04a}).
 Luminosity distances have been derived assuming a flat $\rm
{\Lambda}$-dominated cosmology described by $H_{0} = 71\,\rm{km\,s}^{-1}\,\rm{Mpc}^{-1}$ and $\rm{\Omega_{m}} = 0.27$. 
Observational parameters and results are summarized in Tables~\ref{t1} and \ref{t2}.


\begin{figure*}[htb!]
   \centering
   \includegraphics[width=9cm, angle=-90]{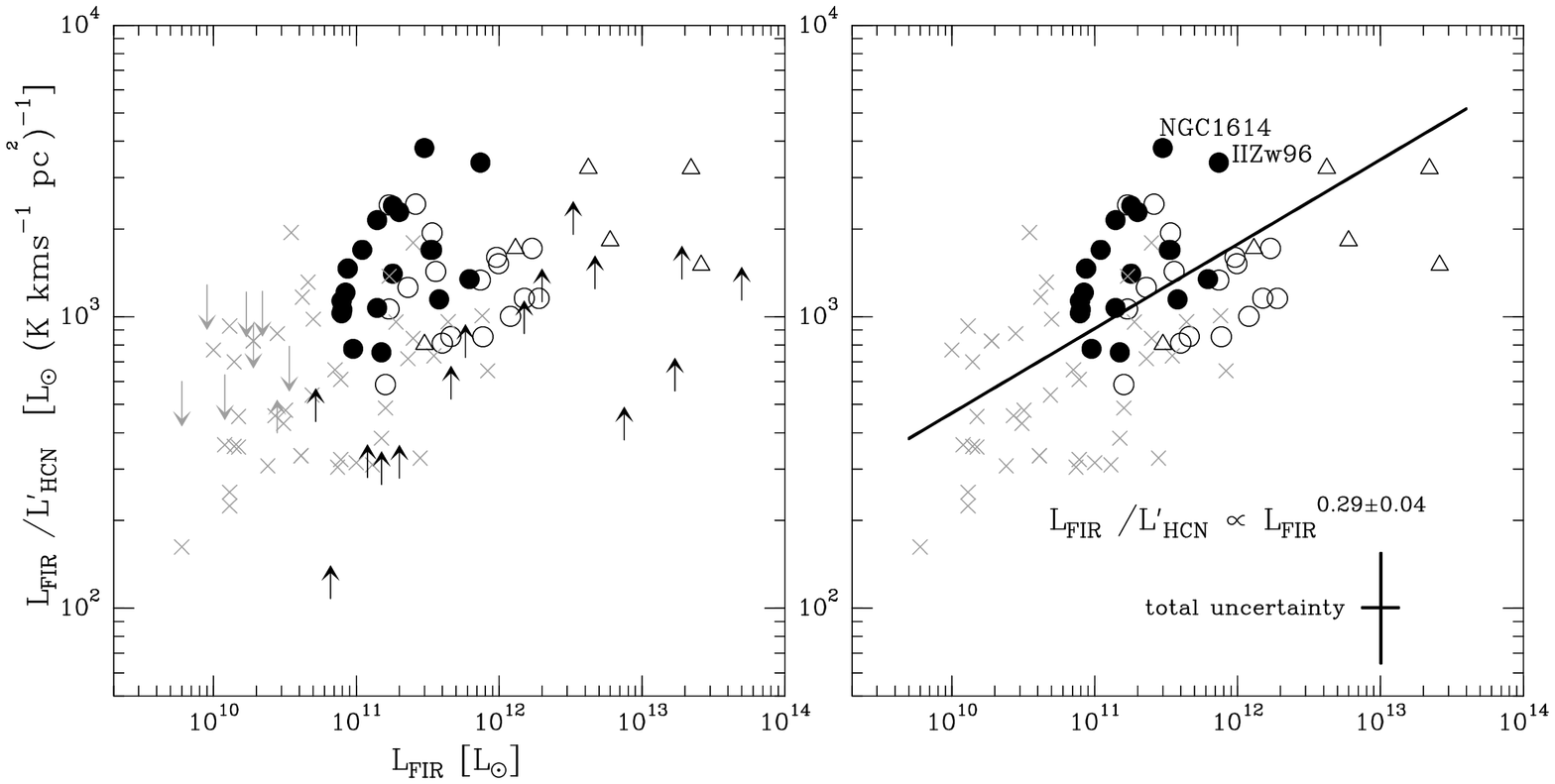}
   \caption{{\bf a)}~({\it left panel}) We plot the $L_{\rm FIR}/L'_{\rm HCN(1-0)}$ luminosity ratio, a proxy for the SFE$_{dense}$ defined in Sect.~\ref{SFE}, as a function of 
   $L_{\rm FIR}$ for different galaxy samples compiled in this work (See Sect.~\ref{Sample} for details). The different symbols represent the normal galaxies and LIRGs/
   ULIRGs published by Gao \& Solomon~(\cite{Gao04a, Gao04b}) (crosses), the LIRGs/ULIRGs published by Graci{\'a}-Carpio et al.~(\cite{Gra08}) (empty circles),   PG 
   QSOs and high-$z$ galaxies published and compiled by Evans et al.~(\cite{Eva06}) and Gao et al.~(\cite{Gao07}), respectively (triangles), and the new sample of LIRGs 
   published in this work (filled circles).   Arrows represent upper and lower limits to $L'_{\rm HCN(1-0)}$.   {\bf b)}~({\it right panel}) Same as {\bf a)}, but limits have not been 
   represented. The solid line visualizes the orthogonal regression fit calculated for the full sample of objects, not taking into account limits.  We highlight the location of NGC~1614 and II~Zw96 in this plot. The total uncertainty of 
   individual data points, including statistical and estimated systematic errors, are indicated by error bars: $\pm$0.13~dex($\pm$30$\%$) in $L_{\rm FIR}$  and $\pm$0.19~dex($\pm$42$\%$) in $L_{\rm FIR}/L'_{\rm HCN(1-0)}$.} 
              \label{SFE}
\end{figure*}

\begin{figure*}[th!]
   \centering
   \includegraphics[width=8.5cm, angle=-90]{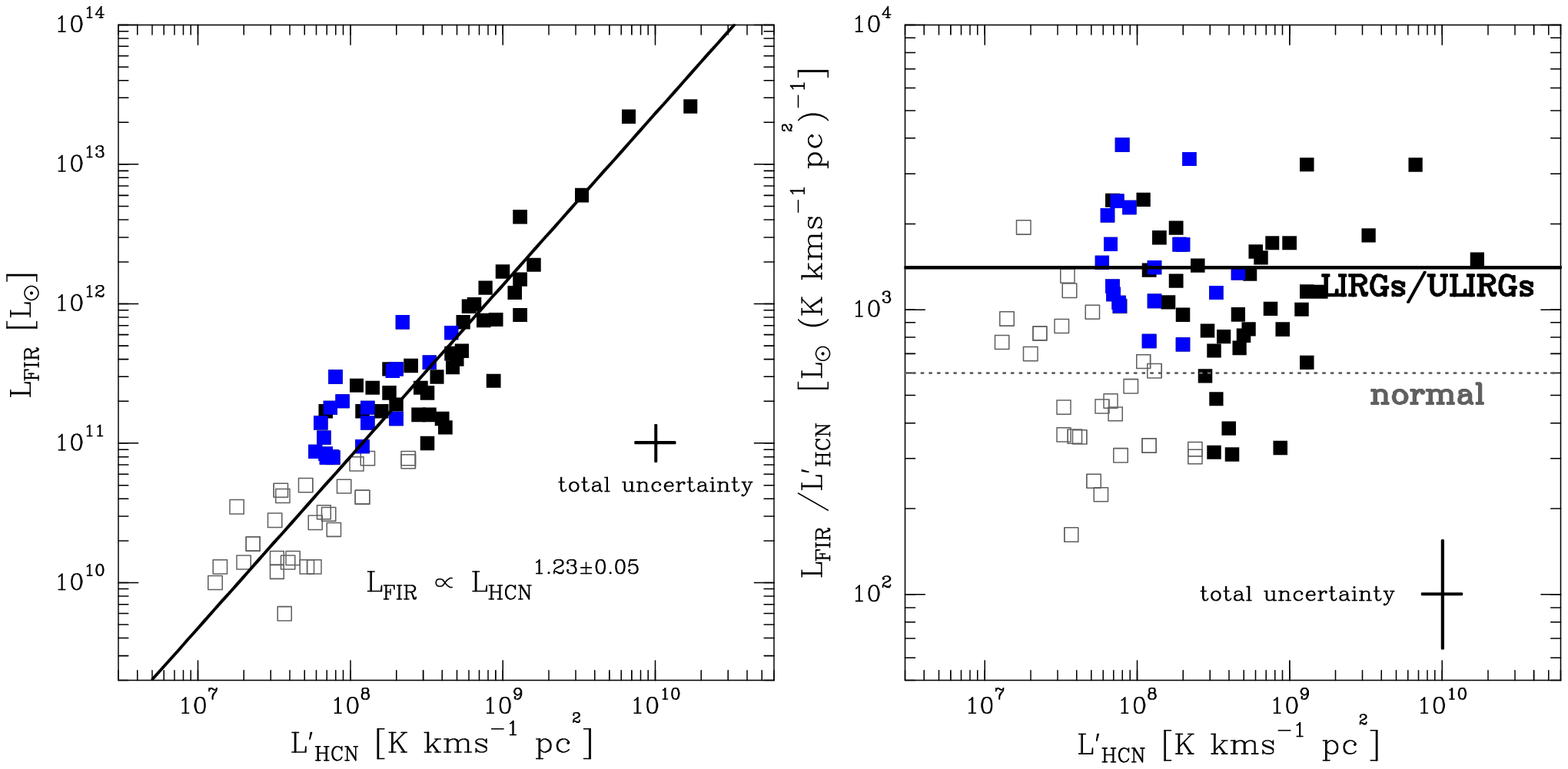}
   \caption{{\bf a)}~({\it left panel}) $L_{\rm FIR}$-$L'_{\rm HCN(1-0)}$ correlation plot derived for different galaxy samples. The solid line visualizes the 
   orthogonal regression fit to the full sample showing a super-linear correlation. The errorbars show the total uncertainty of individual data points 
   ($\pm$0.13~dex=$\pm$30$\%$ in both  $L_{\rm FIR}$ and $L'_{\rm HCN}$) . {\bf b)}~({\it right panel}) We plot the $L_{\rm FIR}/L'_{\rm HCN(1-0)}$ luminosity 
   ratio as a function of $L'_{\rm HCN}$. We display with different symbols normal galaxies  ($L_{\rm IR} < 10^{11} L_{\sun}$, open squares) and luminous infrared galaxies  
   ($L_{\rm IR} > 10^{11} L_{\sun}$, filled squares). The location of the new sample of LIRGs in this diagram is identified by blue color markers. The dashed and continuous 
   horizontal lines  indicate, respectively, the average value of $L_{\rm FIR}/L'_{\rm HCN(1-0)}$ in normal galaxies ($\sim$600$\pm$70$L_{\sun}\,{L'}^{-1}$) and LIRGs/
   ULIRGs ($\sim$1400$\pm$100$L_{\sun}\,{L'}^{-1}$). The errorbars show the total uncertainty of individual data points:  $\pm$0.13~dex($\pm$30$\%$) in $L'_{\rm HCN}$  
   and $\pm$0.19~dex($\pm$42$\%$) in $L_{\rm FIR}/L'_{\rm HCN(1-0)}$.} 
              \label{LFIR-LHCN}
\end{figure*}


\begin{figure}[bth!]
   \centering
   \includegraphics[width=8.5cm, angle=-90]{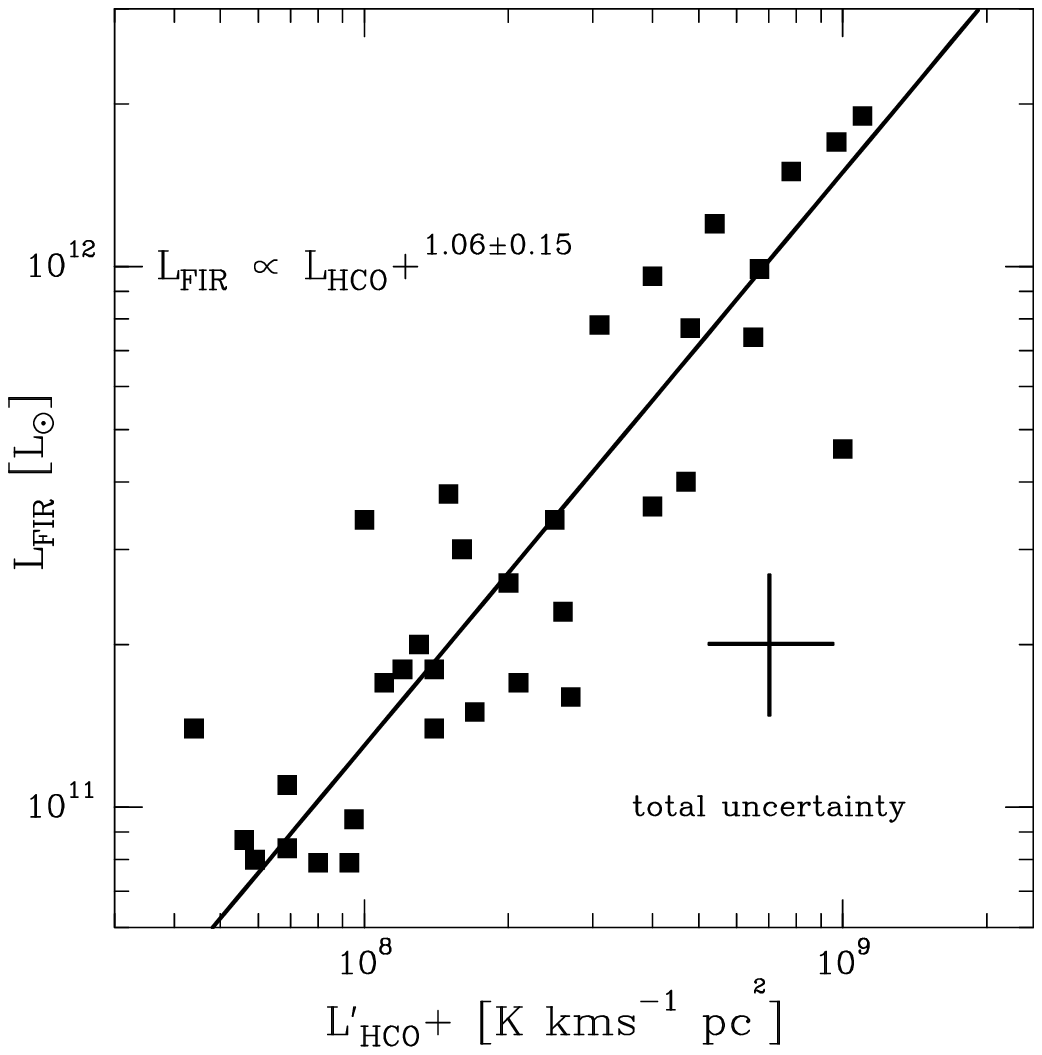}
   \caption{ $L_{\rm FIR}$-$L'_{\rm HCO^+(1-0)}$ correlation plot derived for a subset of LIRGs/ULIRGs. The solid line visualizes the 
   orthogonal regression fit to the sample. Errorbars as in Fig.\ref{LFIR-LHCN}.} 
              \label{hcop}
\end{figure}


\begin{figure}[tbh!]
   \centering
   \includegraphics[width=8cm, angle=-90]{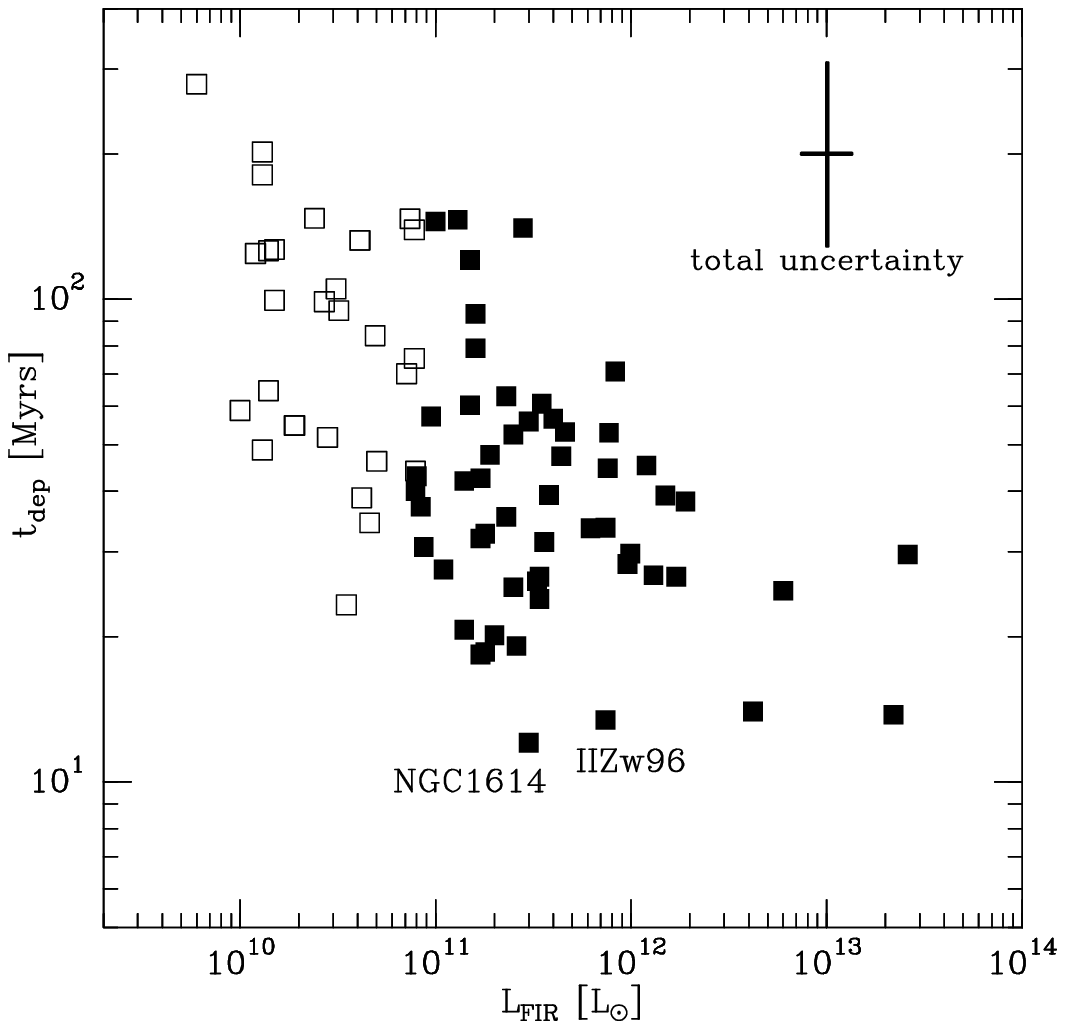}
   \caption{The dense molecular gas depletion or exhaustion time scale, t$_{\rm dep}$$\equiv$1/SFE$_{\rm dense}$ in Myrs, derived for normal galaxies and LIRGs/ULIRGs. 
   Symbols are as in Fig.~\ref{LFIR-LHCN}. We highlight the location of NGC~1614 and II~Zw~96 in the diagram. Errorbars show total uncertainties: $\pm$42$\%$ in 
   t$_{\rm dep}$ and $\pm$30$\%$ in $L_{\rm FIR}$.} 
              \label{Tdepl}
\end{figure}

\section{Star formation laws of the dense gas} \label{SFlaws}

In Sects~\ref{SFElaws} and \ref{KSlaws} we study the star formation efficiency of the dense gas (SFE$_{\rm dense}$) 
and the Kennicutt-Schmidt (KS) laws in the different populations of galaxies compiled for this work, including the new sample of LIRGs.  
Following the same approach adopted by  Graci\'a-Carpio et al.~(\cite{Gra08}),   we derive  
${\rm SFR}\ [M_{\sun}\,{\rm yr^{-1}}]$ from $L^{\rm SFR}_{\rm IR}$ defined as $L^{\rm SFR}_{\rm IR}$=1.3$\times$ $L_{\rm FIR}$(40--500\,$\mu$m). This 
aims at minimizing the possible AGN  contribution to $L_{\rm IR}$(8--1000\,$\mu$m). The AGN contribution to $L_{\rm IR}$ in the population of local
LIRGs with AGN detections is nonetheless known to be small, with a median value of $L_{\rm bol}$(AGN)/$L_{\rm IR}$=0.05 (Alonso-Herrero et al.~\cite{Alo11}). 

In this section we adopt the common assumptions regarding conversion factors used to derive SFR from  $L^{\rm SFR}_{\rm IR}$  and $M_{\rm dense}$ from 
$L'_{\rm HCN(1-0)}$. We first adopt the same universal factor used by Kennicutt~(\cite{Ken98}):

\begin{equation} 
  {\rm SFR}\ [M_{\sun}\,{\rm yr^{-1}}] = 1.7 \times 10^{-10}\,L^{\rm SFR}_{\rm IR}\ [L_{\sun}] 
\end{equation} 

Similarly, $M_{\rm dense}$ is obtained in the following from $L'_{\rm HCN(1-0)}$ assuming a universal conversion factor ($\alpha^{\rm HCN}$) 
similar to that of Gao \& Solomon~(\cite{Gao04a}): 

\begin{equation}
  M_{\rm dense}\ [M_{\sun}] = 10\,L'_{\rm HCN(1-0)}\ [\rm K\,km\,s^{-1}\,pc^{2}] 
\end{equation}

We nevertheless question the grounds for these two working assumptions and the consequences thereof in Sect.~\ref{factors}.

\subsection{Star formation efficiency  (SFE$_{\rm dense}$)}\label{SFElaws}

Figure~\ref{SFE} shows the $L_{\rm FIR}/L'_{\rm HCN(1-0)}$ luminosity ratio versus $L_{\rm FIR}$ for the different populations 
of galaxies in this compilation, defined in Sect.~\ref{Sample}.  We adopt here the common assumption that the 
$L_{\rm FIR}/L'_{\rm HCN(1-0)}$ luminosity ratio is a good proxy for the star formation efficiency of the dense gas 
(SFE$_{\rm dense}$$\equiv$${\rm SFR}/M_{\rm dense} \propto L_{\rm{FIR}}/L'_{\rm{HCN(1-0)}}$). 
We estimate that the total uncertainty on  both $L_{\rm FIR}$ and $L'_{\rm{HCN(1-0)}}$ (and related quantities), with an equal contribution of statistical and 
systematic errors, amounts to $\pm$30$\%$ ($\pm$0.13~dex). The corresponding total uncertainty on $L_{\rm{IR}}/L'_{\rm{HCN(1-0)}}$ is $\pm$42$\%$ ($\pm$0.19~dex).

 In Fig.~\ref{SFE} we can see that SFE$_{\rm dense}$ increases with $L_{\rm FIR}$ from normal galaxies to extreme starbursts, represented 
 by LIRGs, ULIRGs and high-$z$ objects.  In particular, the orthogonal regression fit to the full sample of objects gives a power law index for 
 SFE$_{\rm dense}$ significantly different to zero:

\begin{equation}
  \log{\left( \frac{L_{\rm FIR}}{L'_{\rm HCN(1-0)}} \right)} = (0.29 \pm 0.04)\log{L_{\rm FIR}} - (0.23 \mp 0.42) 
\end{equation} 

\begin{equation} 
  \mathrm{or\ } \frac{L_{\rm FIR}}{L'_{\rm HCN(1-0)}} \simeq 0.59\ L_{\rm FIR}^{0.29} 
\end{equation} 

The location of the new sample of LIRGs in the SFE$_{\rm dense}$ diagrams of Figs. ~\ref{SFE}  and \ref{LFIR-LHCN}b corroborates on a more solid 
statistical basis, compared to the results of Graci\'a-Carpio et al.~(\cite{Gra08}), that SFE$_{\rm dense}$ is on average a factor 
$\sim$2--3 higher in LIRGs/ULIRGs  compared to normal galaxies ($<$SFE$_{\rm dense}$$>$(LIRGs/ULIRGs)
$\sim$1400$\pm$100$\,L_{\sun}\,{L'}^{-1}$;  $<$SFE$_{\rm dense}$$>$(normal)$\sim$600$\pm$70$\,L_{\sun}\,{L'}^{-1}$).  
This difference is a factor of 2 larger than the total uncertainty on individual data points in Fig. ~\ref{SFE}. 
The  total change in SFE$_{\rm dense}$ expands an order of magnitude from normal galaxies to LIRGs/ULIRGs.
Of particular note, 
two  objects of the new sample, NGC~1614 and II~Zw96,  show the highest SFE$_{\rm dense}$ values ($\geq$3500$L_{\sun}\,{L'}^{-1}$) 
thus far reported in a galaxy.  These values are close the upper limit imposed by the maximum 
efficiencies measured in Galactic star-forming cores ($\sim$3900$L_{\sun}\,{L'}^{-1}$; Wu et al.~\cite{Wu05}). The resulting index derived for the  
SFE$_{\rm dense}$ law  (0.29$\pm$0.04) is larger than the one obtained by Graci\'a-Carpio et al.~(\cite{Gra08}) (0.24$ \pm$0.04).

Figure~\ref{LFIR-LHCN}a shows the $L_{\rm FIR}$-$L'_{\rm HCN(1-0)}$ correlation plot obtained for the full sample of galaxies.  
The orthogonal fit to the data points gives a super-linear correlation:

\begin{equation} 
  \log{L_{\rm FIR}} = (1.23 \pm 0.05)\log{L'_{\rm HCN(1-0)}} + (1.07 \mp 0.40) 
\label{hcn}  
\end{equation}

\begin{equation} 
  \mathrm{or\ } L_{\rm FIR} \simeq 12\ {L'}_{\rm HCN(1-0)}^{1.23} 
\end{equation}

The fit is similar to that found by Graci\'a-Carpio et al.~(\cite{Gra08}). With the addition of the new LIRG sample we thus confirm that the  
$L_{\rm FIR}$-$L'_{\rm HCN(1-0)}$ correlation is significantly super-linear. This partly contradicts the first findings of 
Gao \& Solomon~(\cite{Gao04a,Gao04b}) who determined a power index about 1 for the  $L_{\rm FIR}$-$L'_{\rm HCN(1-0)}$ 
scatter plot and a correspondingly constant SFE$_{\rm dense}$ independent of galaxy type. 


Figure~\ref{hcop} shows the $L_{\rm FIR}$-$L'_{\rm HCO^+(1-0)}$ correlation plot obtained for the sample of 34 LIRGs/ULIRGs where
we have obtained data in the HCO$^+$ line\footnote{There is no equivalent survey of HCO$^+$ data in a significant number of normal 
galaxies. The comparison between the star formation properties of the dense molecular in LIRGs/ULIRGs and normal galaxies has thus to rely hereafter mainly on HCN data.} (data from this work and from Graci\'a-Carpio et al.~\cite{Gra06, Gra08}).   
The orthogonal fit to the data points gives a close to linear correlation:

\begin{equation} 
  \log{L_{\rm FIR}} = (1.06 \pm 0.15)\log{L'_{\rm HCO^+(1-0)}} + (2.64 \mp 0.04) 
\end{equation}

\begin{equation} 
  \mathrm{or\ } L_{\rm FIR} \simeq 436\ {L'}_{\rm HCO+(1-0)}^{1.06} 
\end{equation}

 We note that a similar fit using HCN(1--0) data but restricted to the sample of LIRGs/ULIRGs gives also a close to linear power law
($n$=1.10$\pm$0.10). A similar fit of the $L_{\rm FIR}$-$L'_{\rm HCN(1-0)}$ relation restricted to normal galaxies also gives a power index compatible with
unity: $n$=0.95$\pm$0.20. Altogether this is an indication that it is the inclusion of both normal galaxies and LIRGs/ULIRGs in the fit of Eq.~\ref{hcn} that makes the relation between $L_{\rm FIR}$  and $L'_{\rm HCN(1-0)}$ become superlinear. 

 Figure~\ref{Tdepl} shows the dense molecular gas depletion or exhaustion time scale t$_{\rm dep}$ in Myr units, defined as
 t$_{\rm dep}$$\equiv$SFE$_{\rm dense}^{-1}$$\equiv$$M_{\rm dense}\ [M_{\sun}]$/${\rm SFR}\ [M_{\sun}\,{\rm yr^{-1}}] $, as a function of $L_{\rm FIR}$.  NGC~1614 and II~Zw96 lie at the lower end of 
 the t$_{\rm dep}$ distribution and are characterized by significantly short depletion time scales $\simeq$10~Myrs. This is an indication that they 
 represent extreme starburst  systems that will exhaust their dense molecular gas content on time scales comparable to the typical dynamical time scales 
 for these sources (see discussion in Sect.~\ref{Timescales} and in Sect.~3.6 of Alonso-Herrero et al.~\cite{Alo01} for NGC~1614).

 \subsection{Kennicutt-Schmidt power laws}\label{KSlaws}

\begin{figure*}[tbh!]
   \centering
   \includegraphics[width=8cm, angle=-90]{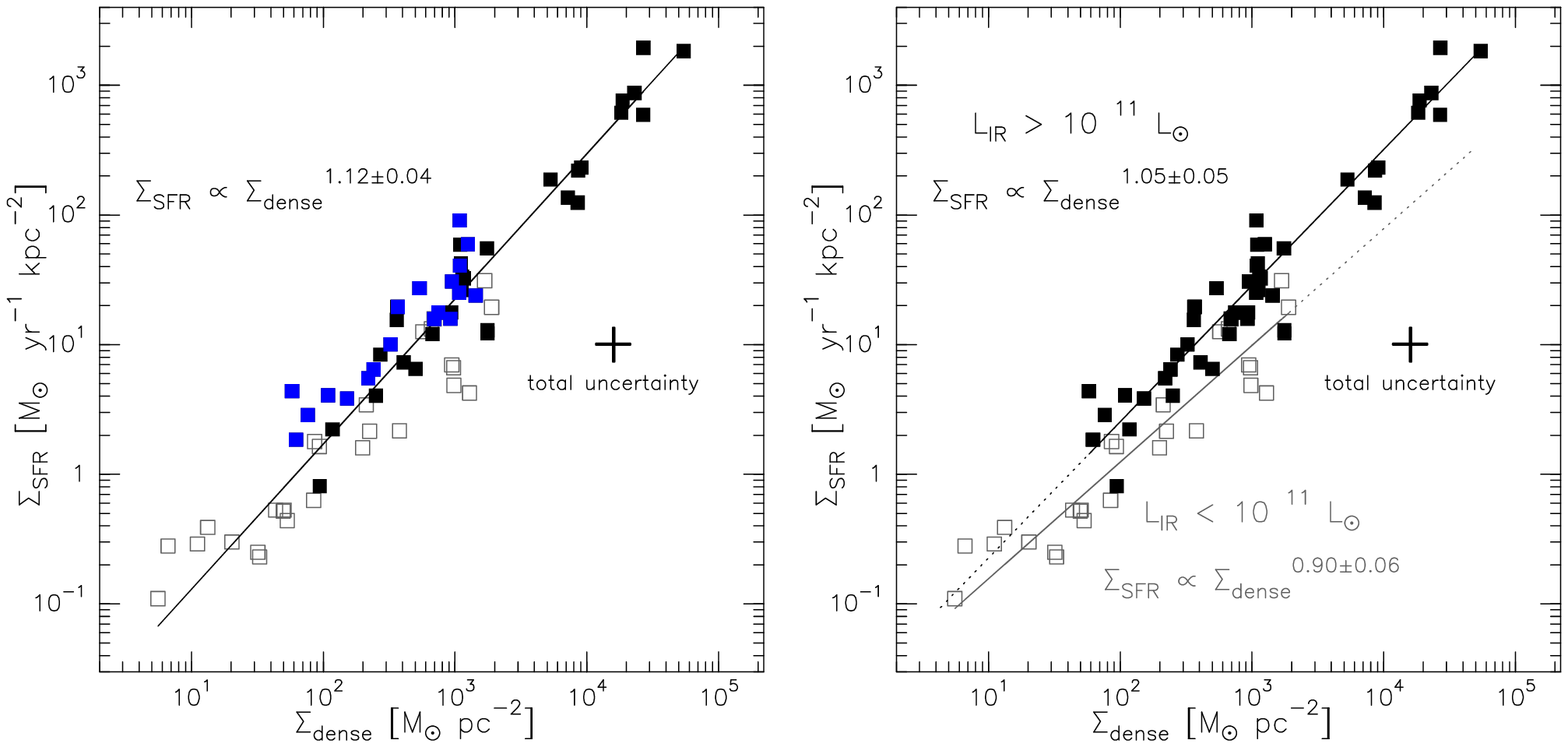}
   \caption{{\bf a)}~({\it left panel}) Star formation rate surface density  $\Sigma_{\rm SFR}$ as a function of dense molecular gas surface 
   density as traced by the HCN(1--0) line, $\Sigma_{\rm dense}$, in different populations of galaxies for which we have an estimate of the 
   molecular disk size. Symbols are as in Fig.~\ref{LFIR-LHCN}. The solid black line is the 
 orthogonal regression fit to the full sample of galaxies.  {\bf b)}~({\it right panel}) Same as {\bf a)}, but showing the two function power law fit   
 to normal galaxies (grey line) and IR luminous galaxies (black line). Here, normal galaxies  ($L_{\rm IR} < 10^{11} L_{\sun}$) are represented by open squares and IR luminous galaxies  ($L_{\rm IR} > 10^{11} L_{\sun}$) by black filled squares as referred to in text. Errorbars show $\pm$30$\%$ uncertainties in $\Sigma_{\rm SFR}$ and $\Sigma_{\rm dense}$.} 
              \label{KS}
\end{figure*}

 Figure~\ref{KS} shows the KS law obtained for the different populations of galaxies compiled for this work., i.e., the
 star formation rate surface density  $\Sigma_{\rm SFR}$ as a function of dense molecular gas surface density as traced by the HCN(1--0) 
 line, $\Sigma_{\rm dense}$. In order to derive $\Sigma_{\rm SFR}$ and $\Sigma_{\rm dense}$ we adopted the molecular gas size estimates  
 from published CO or HCN  interferometer maps, which are available for most of the sources culled from the literature. As for the new sample of LIRGs, we estimated the size of the central star forming region from the high-resolution H$\alpha$ and/or  Pa$\alpha$ images of the galaxies published by 
 Alonso-Herrero et al.~(\cite{Alo06}) and adopted those as representative of the typical sizes of the actively star forming 
 molecular gas complexes.

In Fig.~\ref{KS}a we have represented the derived KS law. The two surface densities, $\Sigma_{\rm SFR}$ and $\Sigma_{\rm dense}$, follow a correlation over more than 4 orders of magnitude in $\Sigma_{\rm dense}$. An orthogonal regression fit to the data results in a KS-law of the dense molecular gas with a power index $N = 1.12 \pm 0.04$:

\begin{equation}
  \log{\Sigma_{\rm SFR}} = (1.12 \pm 0.04)\log{\Sigma_{\rm dense}} + (-2.01 \mp 0.12)
\end{equation}

\begin{equation}
  \mathrm{or\ } \Sigma_{\rm SFR} \simeq 0.010\ \Sigma_{\rm dense}^{1.12}
\end{equation}

The location of the new sample of LIRGs in the KS diagram of Fig. ~\ref{KS}a  helps visualize that extreme starbursts (LIRGs/ULIRGs) and normal galaxies are not fully overlapping in this scatter plot.  In order to quantify if a two-function power law significantly improves the overall fit we have split the sample into normal ($L_{\rm IR} < 10^{11}\,L_{\sun}$) and IR luminous galaxies ($L_{\rm IR} > 10^{11}\,L_{\sun}$). Figure~\ref{KS}b shows the result of the two-function fit, which can be expressed as:

\begin{equation}
  \log{\Sigma_{\rm SFR}} = (0.90 \pm 0.06)\log{\Sigma_{\rm dense}} + (-1.71 \mp 0.15)
\end{equation}

\begin{equation} 
  \mathrm{or\ } \Sigma_{\rm SFR} \simeq 0.02\ \Sigma_{\rm dense}^{0.90}
\end{equation}

\noindent for galaxies with $L_{\rm IR} < 10^{11}\,L_{\sun}$, and:

\begin{equation}
  \log{\Sigma_{\rm SFR}} = (1.05 \pm 0.05)\log{\Sigma_{\rm dense}} + (-1.70\mp 0.16)
\end{equation}

\begin{equation} 
  \mathrm{or\ } \Sigma_{\rm SFR} \simeq 0.02\ \Sigma_{\rm dense}^{1.05}
\end{equation}

\noindent for local and high-$z$ IR luminous galaxies with $L_{\rm IR} \geq 10^{11}\,L_{\sun}$.

Whereas the index of the power law is not substantially different in normal galaxies ($N = 0.90 \pm 0.06$) and 
LIRGs/ULIRGs ($N = 1.05 \pm 0.05$)  this fit reinforces the idea of duality hinted at by the solution that  Graci\'a-Carpio et al.~(\cite{Gra08}) 
found using a smaller sample of LIRGs/ULIRGs. As illustrated in  Fig.~\ref{KS}b, the extrapolation 
of the KS law fitting normal galaxies to the highest values of $\Sigma_{\rm dense}$ ($\sim$a few10$^4$M$_{\sun}$pc$^{-2}$) underpredicts the  $\Sigma_{\rm SFR}$ in 
IR luminous galaxies by up to a factor $\sim$5. This is a factor $\sim$4 larger than the total uncertainty on $\Sigma_{\rm SFR}$. Within the range of gas surface densities shared by normal galaxies and LIRGs/ULIRGs, $\Sigma_{\rm dense}$$\sim$10$^2$--2 10$^3$M$_{\sun}$pc$^{-2}$, the disagreement between the two laws stays within a factor 2 to 3, i.e., still a factor $\sim$2--2.5 larger than the total uncertainty on $\Sigma_{\rm SFR}$.


We have evaluated the goodness of the two-function power law fit  by a standard $\chi^{2}$ analysis. We find that $\chi^{2}$ decreases by a factor of 1.4 in the dual fit compared to the single power law, an indication that the dual KS law for the dense gas qualifies as a better description of the data. This result is reminiscent of the bimodality found in the star formation laws derived in normal star forming galaxies and mergers (LIRGs, ULIRGs and SMGs) from CO line data (Genzel et al.~\cite{Gen10}; Daddi et al.~\cite{Dad10}). We nevertheless analyse in Sect.~\ref{stat-tests} what are the potential biases of the two-function fit of Eqs.~(11--14), motivated by the division of our sample at $L_{\rm IR}=10^{11}\,L_{\sun}$.

   In our derivation of KS laws we have used different methods to derive the typical sizes assumed to be common for the  actively star forming regions and the dense molecular gas disks. This is certainly a source of uncertainty, already included in our error budget. We nevertheless note that if there is any bias it will affect both the estimated  $\Sigma_{\rm SFR}$ and $\Sigma_{\rm dense}$ to the same extent. These uncertainties would make all data points to be shifted along trajectories that correspond to straight lines of slope unity. The fits found for the KS laws are very close to the linear behavior and the described uncertainties 
  associated to source sizes cannot thus create a {\it fake} segregation. 
  
%


\begin{figure*}[th!]
\centering
\includegraphics[width=0.33\hsize]{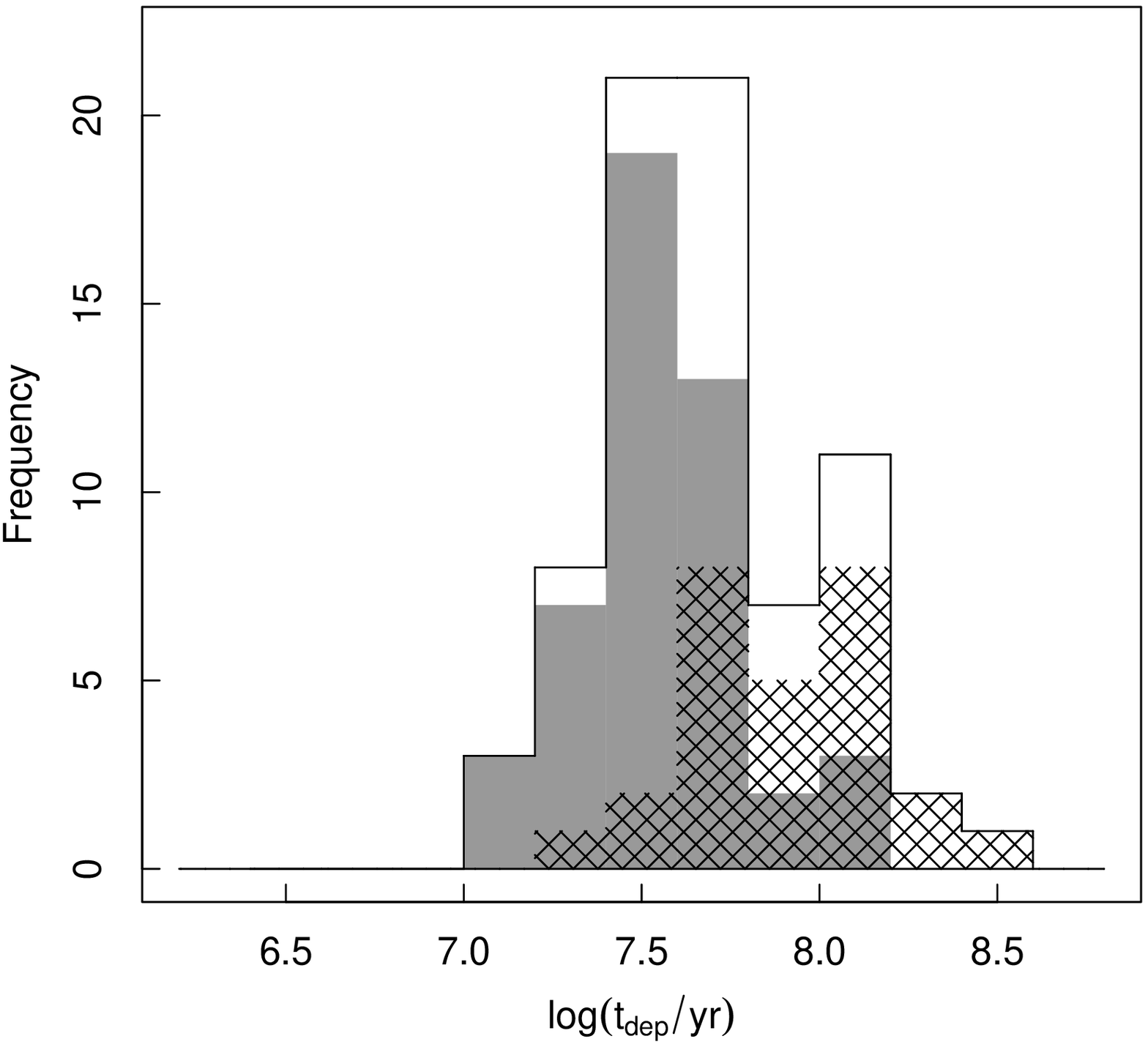}
\includegraphics[width=0.33\hsize]{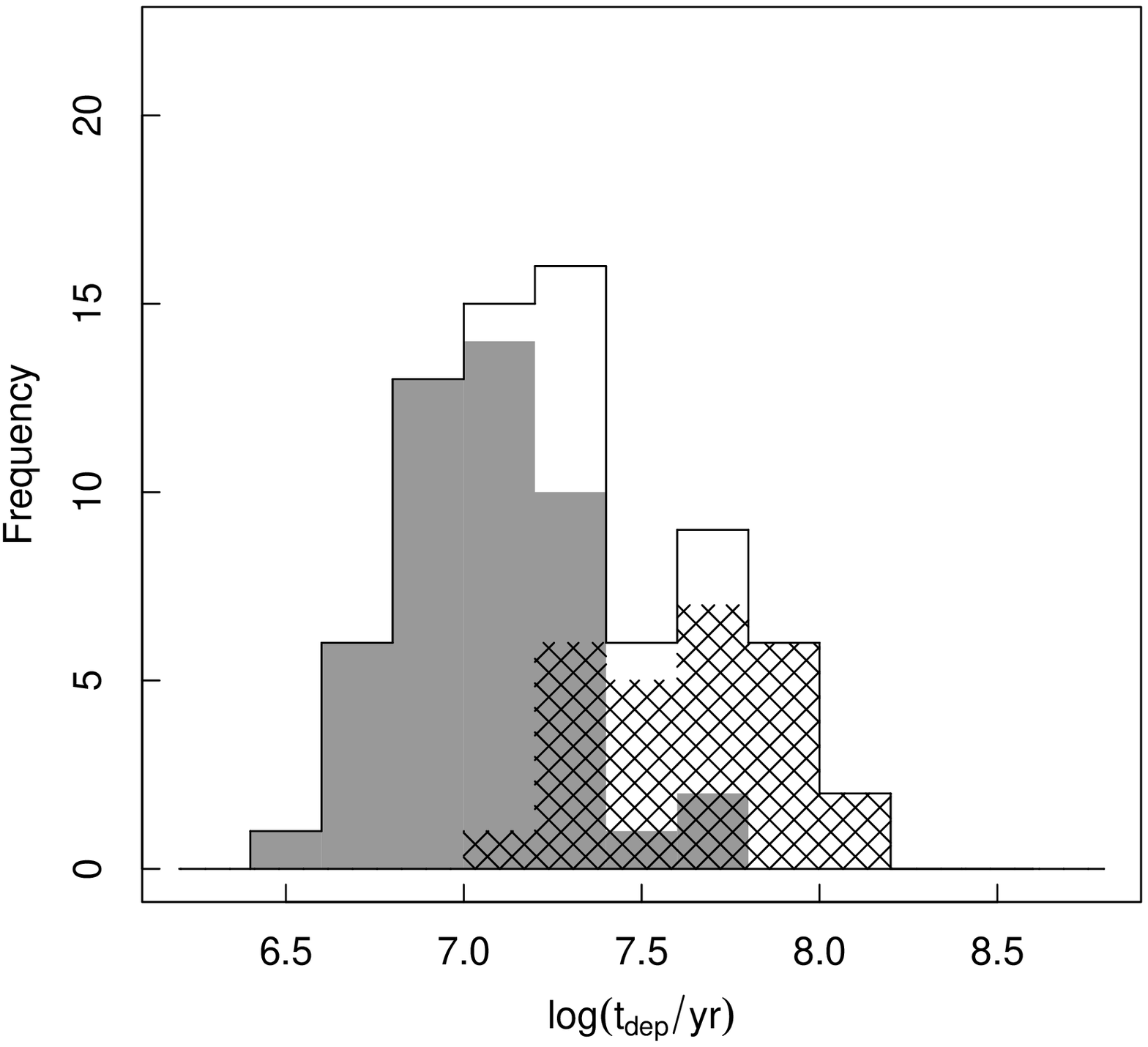}
\includegraphics[width=0.33\hsize]{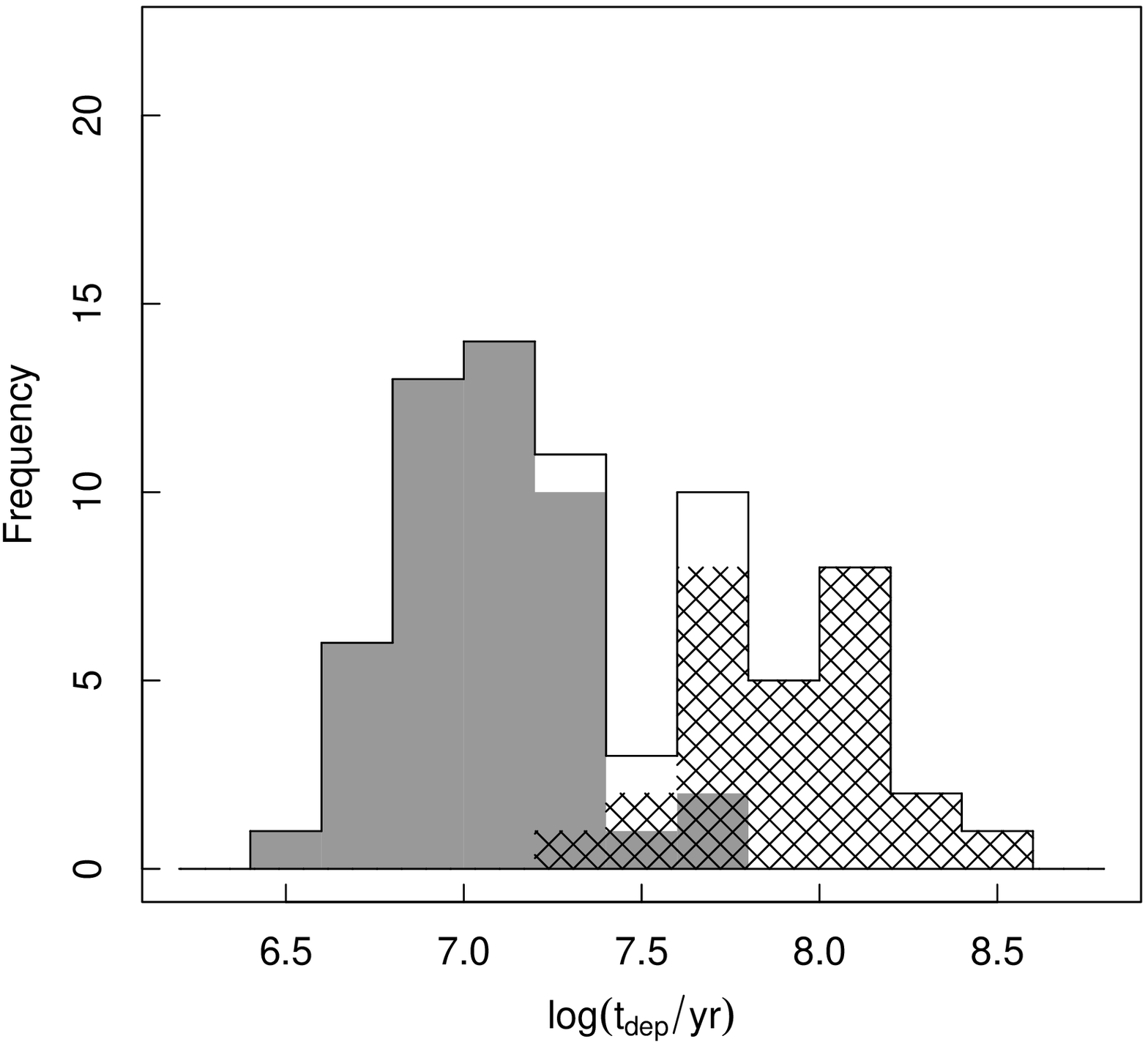}
\caption{Histograms of depletion times for the dense gas in the whole sample of galaxies (solid line) and in the populations of LIRGs/ULIRGs (grey filled) and of normal galaxies (hatched). 
The depletion times in the left-hand panel are calculated assuming the same FIR-to-SFR and HCN-to-mass conversion factors for the two populations. In the middle panel, the two conversion factors are modified (see Sect.~\ref{factors}), whereas in the right-hand panel only the  HCN-to-mass  factor is changed.
} 
\label{histo-dip}
\end{figure*}


\section{Statistical assessment of the two-function fit} \label{stat-tests}

The results found in the previous section suggest that the KS law for the dense gas in  LIRGs/ULIRGs differs from that in normal star-forming galaxies. In this section we quantify this conclusion by addressing the statistical biases inherent to these results.

\subsection{Bimodality}

We first analyse if the star formation laws derived from HCN qualify as bimodal. The definition of bimodality is very strict as it requires the existence of two distinct populations within a given sample  with little if any object connecting the two populations, reflecting two different modes. We can assess if the star formation laws derived from HCN qualify as strictly bimodal by inspecting the distribution of depletion times. Fig.~\ref{histo-dip} shows  histograms of $t_\mathrm{dep}$ in the whole sample and in the groups of LIRGs/ULIRGs and of normal galaxies derived using different approaches as to the assumed conversion factors. As expected from the difference in the mean values derived in Sect.~\ref{SFlaws}, LIRGs/ULIRGs and  normal galaxies tend to populate different regions of the joint histogram. This yields the impression of a bimodal distribution with two unequally populated peaks. Assuming the non-universal conversion factors discussed in Sect.~\ref{factors} certainly increases the spread in $t_\mathrm{dep}$ between normal galaxies and LIRGs/ULIRGs (Fig.~\ref{histo-dip}).

Hartigan \& Hartigan~(\cite{Har85}) proposed a test that measures the goodness of {\it unimodality} as the best description of a statistical sample. The test measures the {\it dip} of a sample, defined as the maximum distance between a given empirical distribution and its corresponding best fitting unimodal distribution. For a given sample size, the deeper the dip, the less likely is that the distribution is unimodal. An updated version of the Hartigan dip test is implemented in the R package, a free software environment for statistical computing \footnote{http://www.r-project.org/}.  We have applied it to our sample of galaxies, 
in order to evaluate if the set of depletion times is compatible with an 
unimodal distribution. The test indicates that the dip of an unimodal 
distribution could be as large as measured for the depletion times in $\sim$90$\%$ of the cases, which otherwise implies that unimodality is far from being rejected. These admittedly high probabilities decrease to $\sim$80$\%$ if we assume the revised conversion factors for $\Sigma_{\rm SFR}$ and $\Sigma_{\rm dense}$ discussed in Sect.~\ref{factors}, and could reach $\sim$20$\%$ if we only changed the conversion factor for $\Sigma_{\rm dense}$, i.e., if we adopted a similar approach to that of Genzel et al.~(\cite{Gen10}).

We nevertheless conclude that we cannot claim that the star formation process is strictly bimodal on the basis of the depletion timescale distribution. This conclusion holds specially in the case where universal conversion factors are adopted. 



\begin{figure*}[th!]
\centering
\includegraphics[width=0.40\hsize]{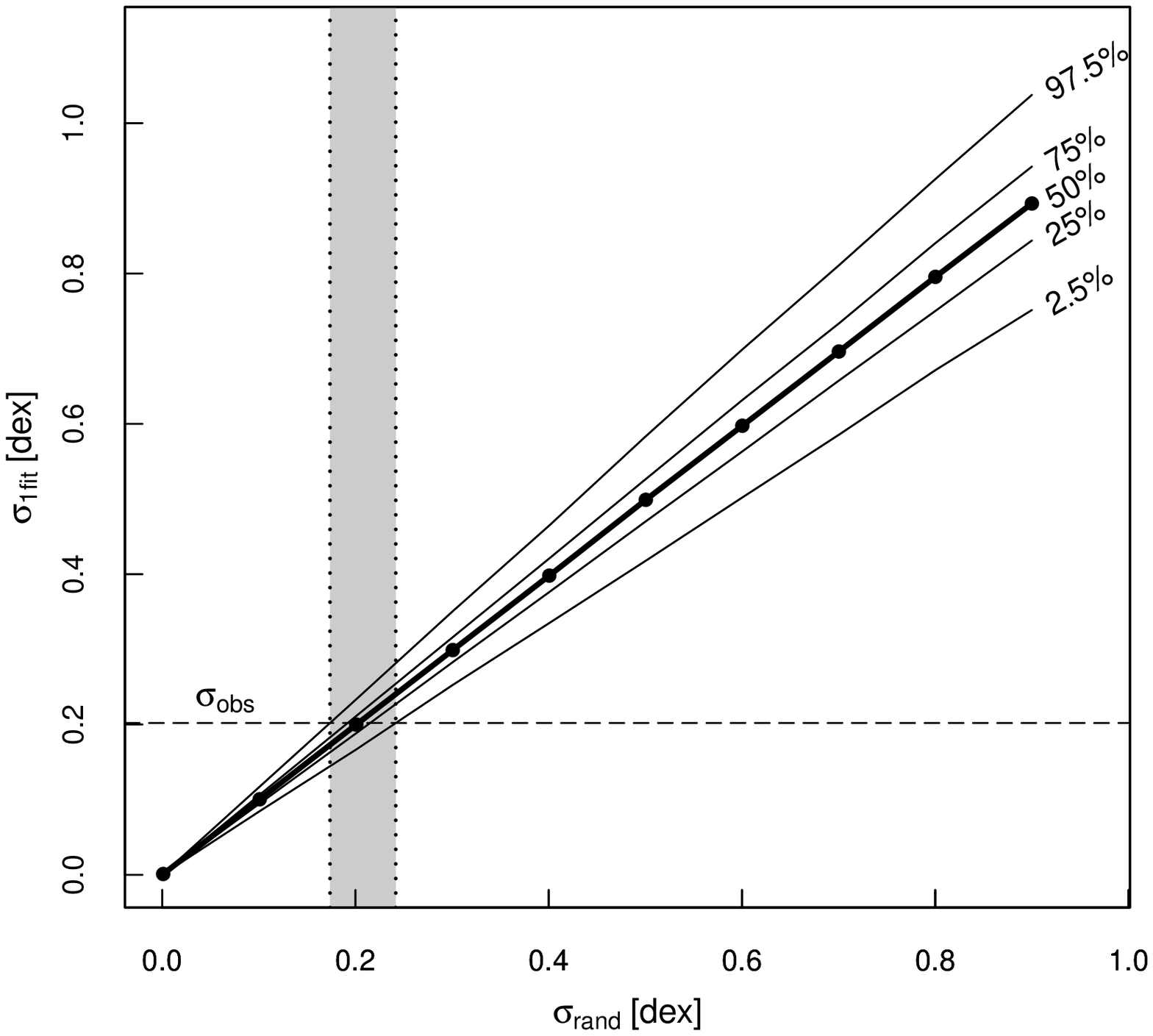}
\includegraphics[width=0.40\hsize]{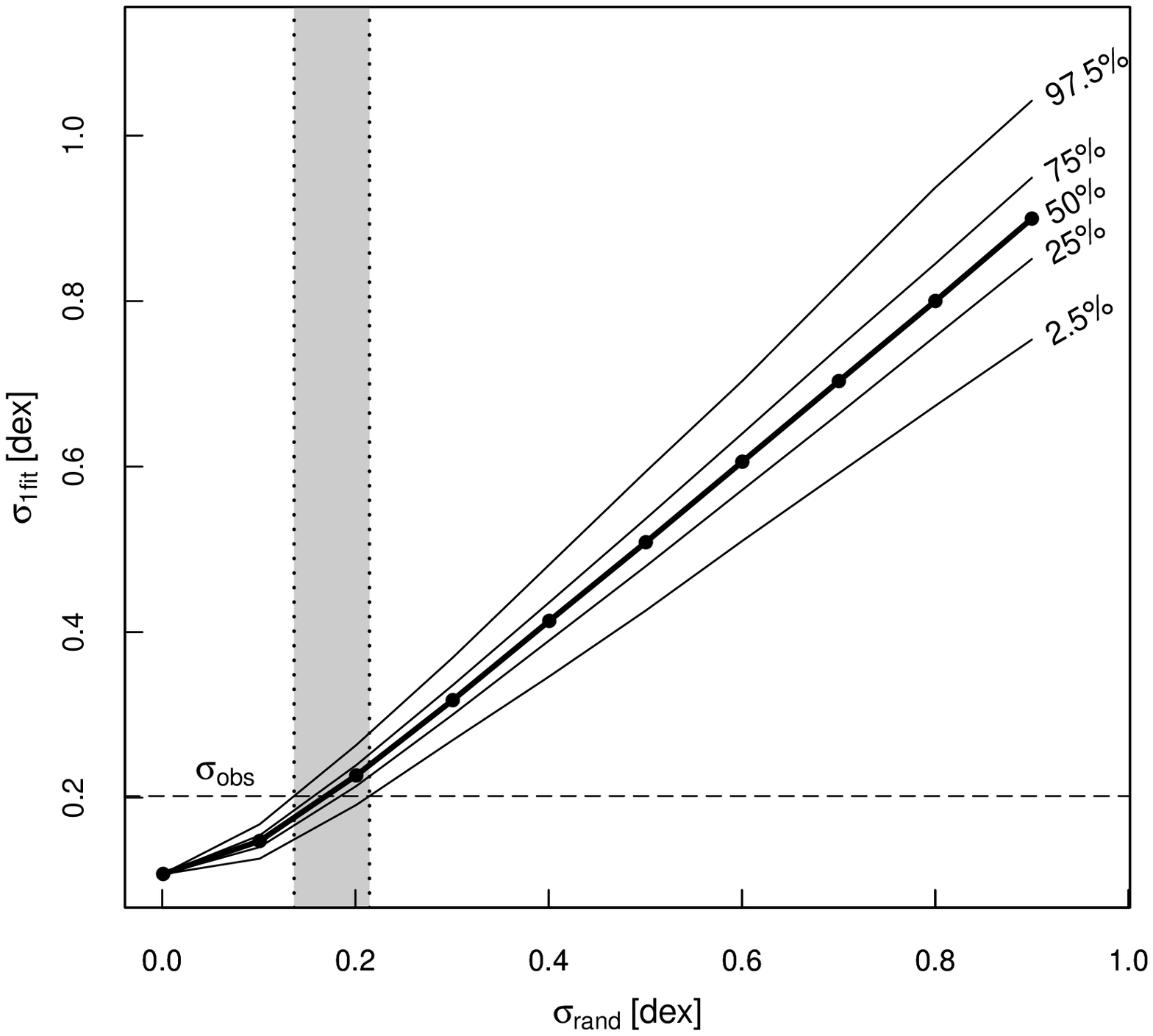}
\\
\includegraphics[width=0.40\hsize]{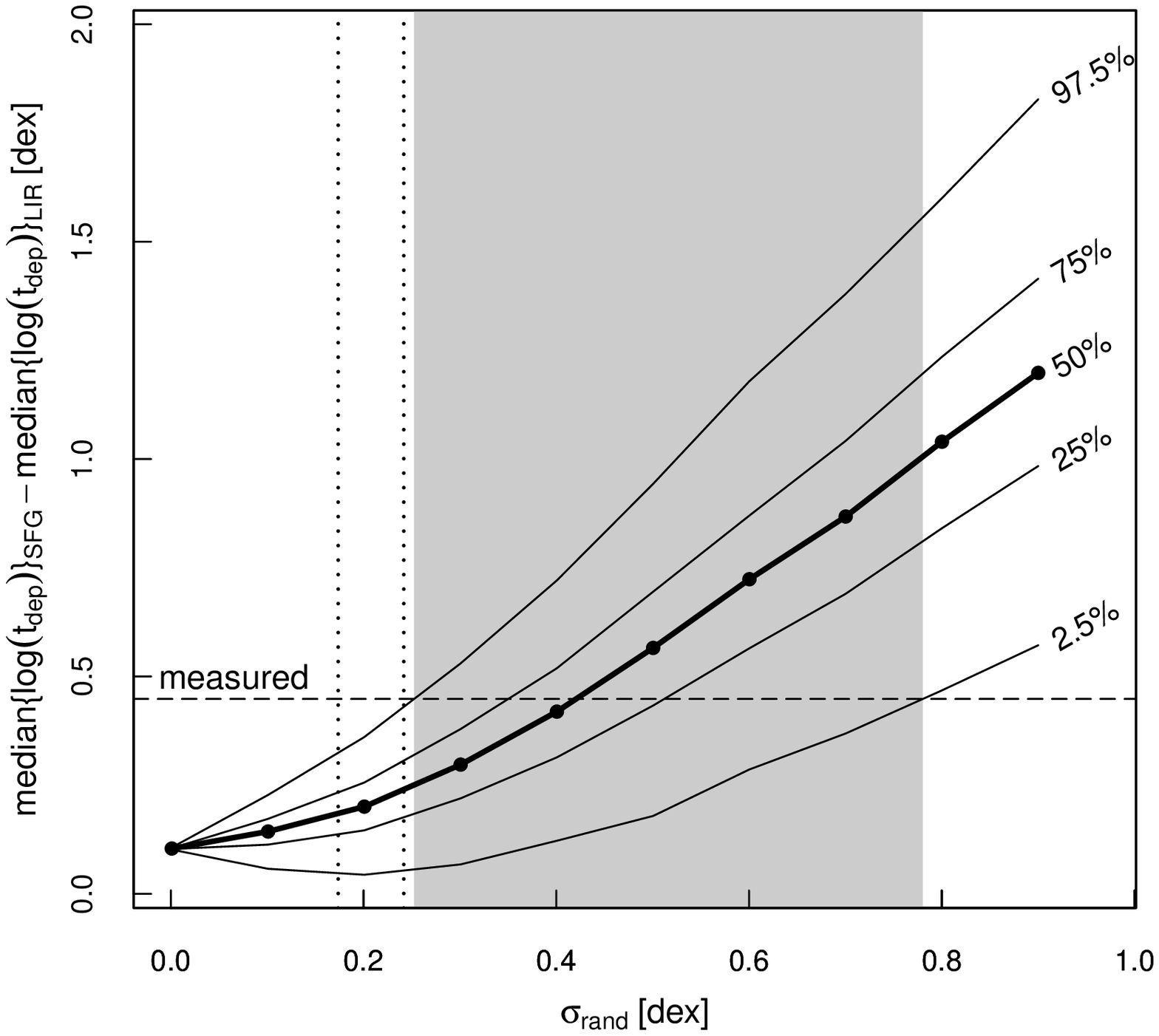}
\includegraphics[width=0.40\hsize]{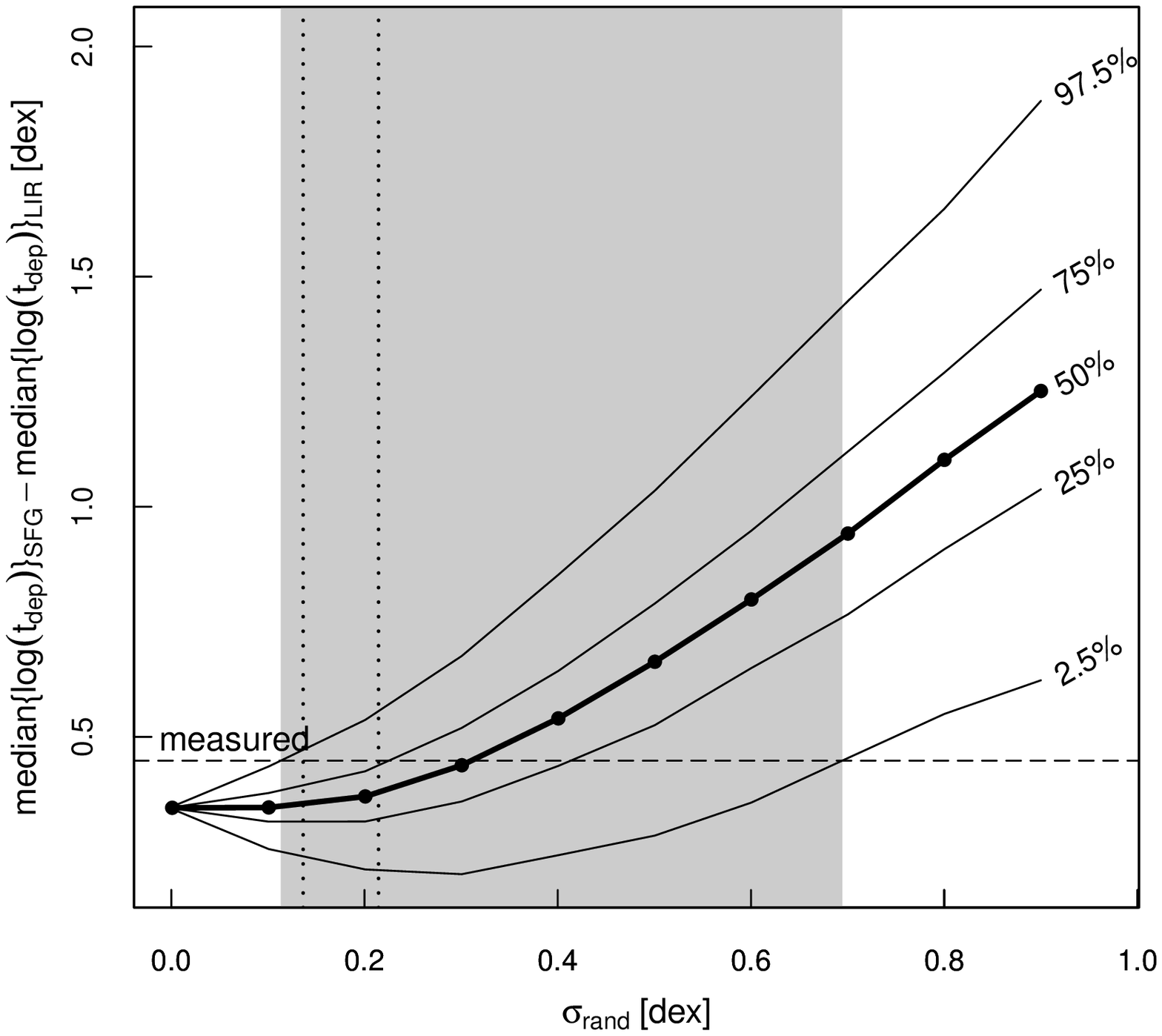}
\\
\includegraphics[width=0.40\hsize]{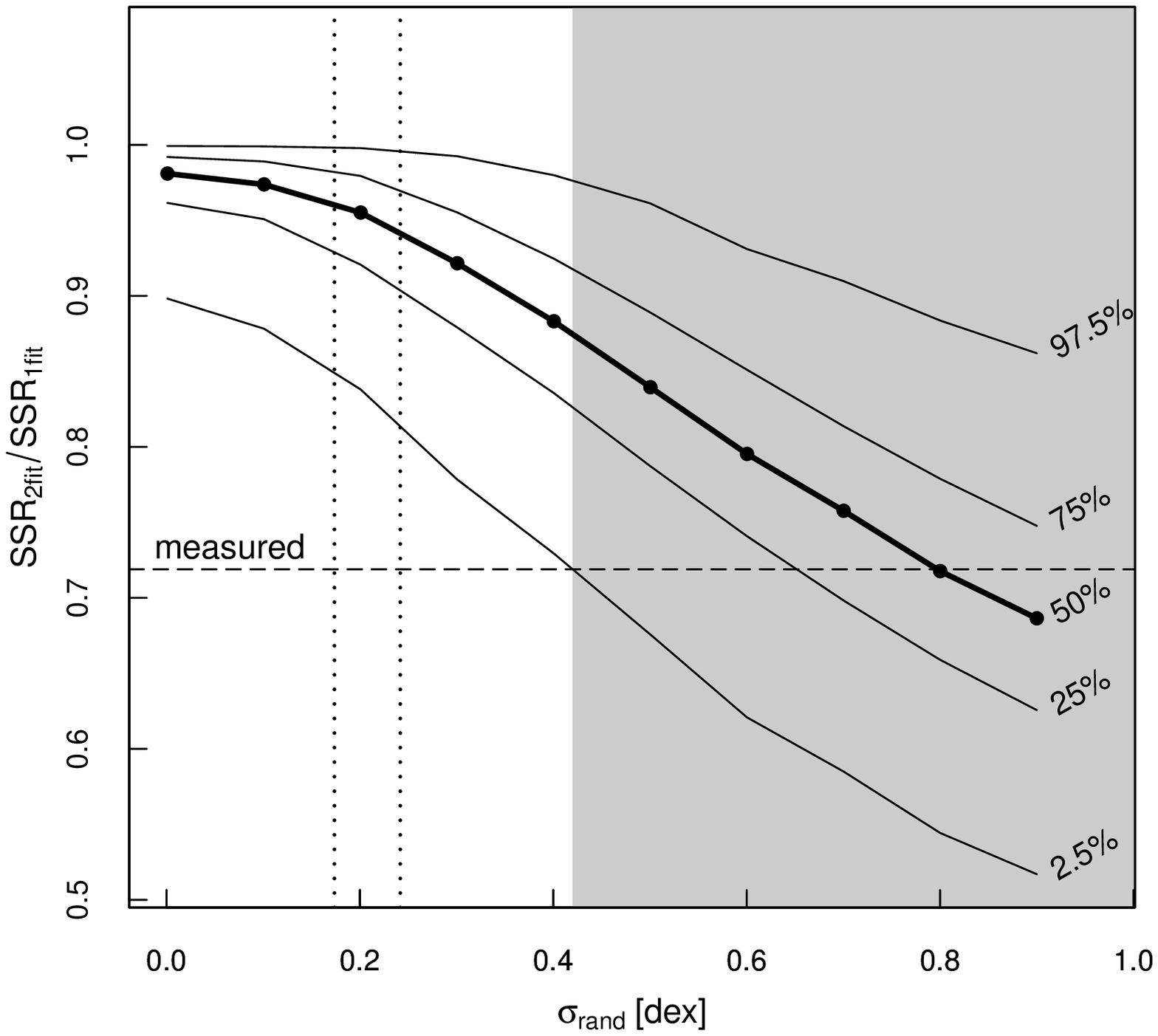}
\includegraphics[width=0.40\hsize]{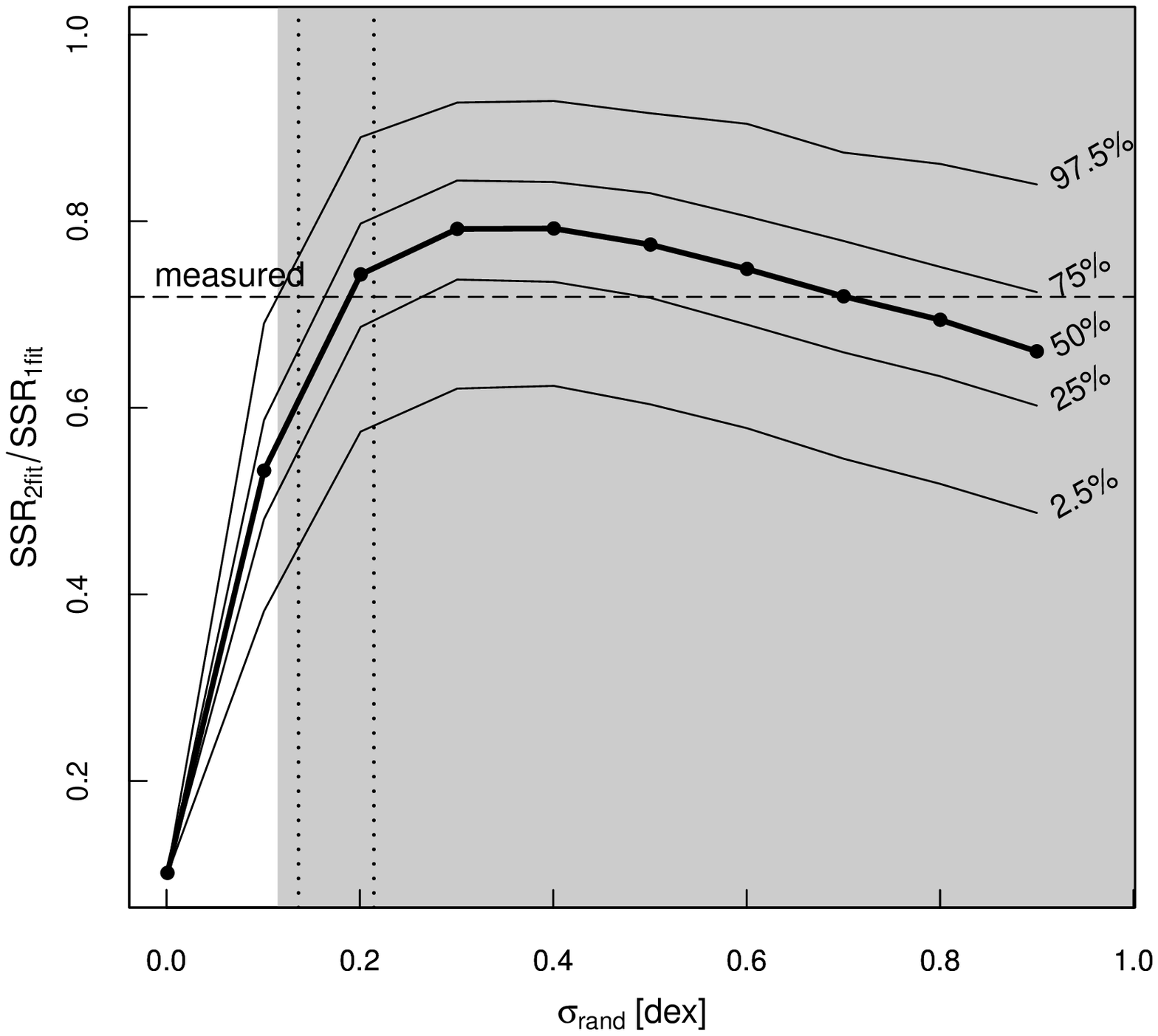}

\caption{Results from testing the null hypotheses of a single KS law (left column) and of a double KS law (right column). Each panel shows the $2.5\%, 25\%, 50\%, 75\%$ and $97.5\%$ percentiles  of $\sigma_\mathrm{1fit}$  (top), $\Delta\log($t$_\mathrm{dep})$ (middle) and $SSR_\mathrm{2fit}/SSR_\mathrm{1fit}$ (bottom) as a function of $\sigma_\mathrm{rand}$. The measured values are indicated by dashed horizontal lines. A gray rectangle represents the $95\%$ confidence interval for $\sigma_\mathrm{rand}$ derived in each panel from the measured value. Dotted lines show the confidence intervals derived from $\sigma_\mathrm{1fit}$ (top panels).}
\label{scatter}
\end{figure*}


\subsection{Universality of the KS law for the dense gas}

The fact that the two-function fit is a significant improvement over the one-function fit suggests nevertheless the existence of an underlying duality in the distribution. Due to our {\it a priori} arbitrary splitting of the galaxy sample into two groups, defined by the dividing line at $L_{\rm IR}=10^{11}\,L_{\sun}$, the evaluation of 
average galaxy properties in the two sub-samples is subject to potential biases, however. The underlying reason is that the selection cut used to separate the populations is not orthogonal to the variables being fit. 

In the following we test the null hypothesis that a single KS law with a fixed power law index for both LIRGs/ULIRGs and normal galaxies (e.g., Eq.~9) is a suitable representation of our data. This hypothesis can be rejected if, on the assumption that it holds true, 
the measured values of three chosen observables are unlikely to result from pure random scatter. The observables used as constraints are:

\begin{itemize}
\item The scatter of the data  with respect to the one-function power law fitted to the data (Eq.~9): $\sigma_\mathrm{1fit}\equiv\sqrt{SSR_\mathrm{1fit}/(\mathcal{N}-2)}$. $SSR_\mathrm{1fit}$ is the sum of squared residuals and $\mathcal{N}$ is the number of sources. The observable $\sigma_\mathrm{1fit}$ tells us the amount of random scatter we can plausibly assume. The value derived for $\sigma_\mathrm{1fit}$ is $0.20$~dex.

\item The difference, in median, between the logarithms of the depletion times of the two populations: $\Delta\log($t$_\mathrm{dep})\equiv\mathrm{median}\{\log($t$_\mathrm{dep})\}_\mathrm{normal}-\mathrm{median}\{\log($t$_\mathrm{dep})\}_\mathrm{(U)LIRG}$. The measured value for $\Delta\log($t$_\mathrm{dep})$ is $0.45$~dex.
The parameter $\Delta\log($t$_\mathrm{dep})$  measures the difference between LIRGs/ULIRGs and normal galaxies, although it is biased towards positive values. The reason is that $\Sigma_\mathrm{SFR}$ is essentially proportional to $L_\mathrm{IR}$, so that a positive noise-driven error in the  $L_\mathrm{IR}$ of a galaxy simultaneously underestimates its t$_\mathrm{dep}$  and increases the probability that the object is classified as LIRG/ULIRG. Superlinear KS laws as that of Eq.~(9) (slope$\sim1.12$) enhance this bias. 

\item  The ratio of  $SSR_\mathrm{2fit}/SSR_\mathrm{1fit}$, where $SSR_\mathrm{2fit}$  is the sum of squared residuals when using two functions to fit the data as in Sect.~\ref{KSlaws}. The measured value for this ratio is about 0.70. This ratio measures the improvement in the fit when separate KS laws are adopted for the two populations. $SSR_\mathrm{2fit}/SSR_\mathrm{1fit}$ is always smaller or equal than one, since a two function fit can always match or even improve the solution of the one function fit.   
\end{itemize}

In order to calculate the corresponding probability distributions of the observables we proceed as follows: 

 \begin{itemize}

\item We assume that the intrinsic distribution of data in the $(\log\Sigma_\mathrm{dense},\log\Sigma_\mathrm{SFR})$ plane is given by the corrected values that result from the orthogonal least squares fit to our observations derived in Sect.~\ref{KSlaws}. These points follow Eq.~(9) by construction. 

\item We can evaluate the effect of the scatter of the distribution on the definition of the subsamples by scrambling our sample independently along the $\Sigma_{\rm dense}$ and $\Sigma_{\rm SFR}$ axes with a Gaussian scatter, hereafter denoted  as $\sigma_\mathrm{rand}$.  The galaxies are then re-classified either as LIRG/ULIRG or as normal after rescaling 
 the IR luminosity of each object by the randomized-to-observed ratio of $\Sigma_\mathrm{SFR}$. This mimics the effect of errors on the classification.  

\item The scatter parameter $\sigma_\mathrm{rand}$ appears as a free parameter in our analysis, since there may be sources of random scatter other than the observational error (0.13 dex). We take 10 values of $\sigma_\mathrm{rand}$ from $\sim0$ dex to $0.9$~dex and perform $5\times10^3$ randomizations for each value. Each randomized set of data is analysed as the observed data to derive the probability distributions of our observables as a function of $\sigma_\mathrm{rand}$.


\end{itemize}

The left column of Fig.~\ref{scatter} shows the $2.5\%, 25\%, 50\%, 75\%$ and $97.5\%$ percentiles  of $\sigma_\mathrm{1fit}$  (top), $\Delta\log($t$_\mathrm{dep})$ (middle) and $SSR_\mathrm{2fit}/SSR_\mathrm{1fit}$ (bottom) as a function of $\sigma_\mathrm{rand}$ assuming the null hypothesis. 
From each panel we can derive a $95\%$ confidence interval for $\sigma_\mathrm{rand}$   (gray bands in the plots) 
by requiring that the probability to get the measured value lies within the $2.5\%-97.5\%$ range.
As shown in the top panel, $\sigma_\mathrm{rand}$ should lie in the range $0.18-0.24$~dex to be compatible with our measurement of  $\sigma_\mathrm{1fit}$ at the $95\%$ confidence level. Outside that range, $\sigma_\mathrm{1fit}$ would be either smaller or larger than measured (0.20~dex) in $97.5\%$ of the cases. Comparing the three panels, we clearly see that this confidence interval is formally incompatible with those derived from the measured  $\Delta\log($t$_\mathrm{dep})$ ($0.25-0.81$~dex) and $SSR_\mathrm{2fit}/SSR_\mathrm{1fit}$ ($>0.42$~dex).

We can thus conclude that the universal one-function fit is not the {\it best} description of the data. Although the ($\Sigma_{\rm SFR}$, $\Sigma_{\rm dense}$) distribution does not qualify as strictly bimodal according to the dip test, the two-function fit to the data qualifies as a better description of the star formation laws derived from HCN. None of the biases related to the intrinsic non-linearity of the law 
or to the a priori arbitrary cut in $L_{\rm IR}$ to divide the sample are able to account for the observed differences between normal galaxies and LIRGs/ULIRGs discussed in Sects.~\ref{SFElaws} and \ref{KSlaws}.

Furthermore, following a similar proof by contradiction approach, we have carried out a similar test assuming a double KS law determined by the best two-function fit of Eqs.~(11--14) discussed in Sect.~\ref{KSlaws}. Equivalent plots to those discussed above are shown in the right column of Fig.~\ref{scatter}. These plots clearly show that the hypothesis of a double KS law is formally in agreement with the measured values. Therefore, although the test cannot prove that the hypothesis of a double KS law is the only solution, we can  conclude that it is a better description of the star formation laws derived from HCN than a single KS law. We note that it could be possible to save the single KS law model by assuming that a significant fraction of the measured scatter is not random, but due to the existence of {\em hidden} variables relevant to the star-formation process like global dynamical 
time scales. This alternative is explored in Sect.~\ref{Timescales}.

\section{Conversion factors and star formation laws}\label{factors}

We discuss in Sect.~\ref{ratios} the validity of the  $M_{\rm dense}$ -- $L'_{\rm HCN(1-0)}$ conversion factor adopted above for LIRGs/ULIRGs  and consider in  Sect.~\ref{SFR} the biases in estimating SFR for normal galaxies from  $L^{\rm SFR}_{\rm IR}$.  The consequences of revising conversion factors for the star formation laws obtained are discussed in Sect.~\ref{SFlaws-rev}.

\subsection{The HCN--to--M$_{\rm dense}$ conversion factor in LIRGs/ULIRGs} \label{ratios}

It is currently accepted that the conversion factor between the CO luminosity ($L'_{\rm CO(1-0)}$) and the mass of 
molecular gas ($M_{\rm gas}$) is lower in mergers than in the Milky Way (MW)  ($\alpha^{\rm CO}_{\rm MW}$=4.8$M_{\sun}\,{L'}^{-1}$). 
The first conclusive evidence of a different conversion factor in mergers was found by Downes \& Solomon~(\cite{Dow98}). 
These authors derived  that the $M_{\rm gas}$ -- $L'_{\rm CO(1-0)}$ conversion factor in ULIRGs is $\sim$1/5 of the MW's value 
(this lowering factor is 1/10 in the case of Arp220).  More generally, the analysis of CO line emission by radiative transfer models indicates that 
the physical conditions found in the CO clouds of mergers and nuclear starbursts 
(densities n(H$_2$)$\simeq$10$^3$cm$^{-3}$-10$^4$cm$^{-3}$ and  brightness temperatures T$_{\rm R}$$\simeq$20--50~K) imply that 
the CO conversion factor, which scales as  n(H$_2$)$^{1/2}$/T$_{\rm R}$,  is typically about (1/2 --1/4)  
$\times$  $\alpha^{\rm CO}_{MW}$ (Solomon et al~\cite{Sol97}; Scoville et al.~\cite{Sco97}).  A clear-cut argument in favor of a lower CO 
conversion factor in mergers is found in the fact that the canonical MW value of $\alpha^{\rm CO}$ applied to  $L'_{\rm CO(1-0)}$ predicts gas 
masses that are larger than the dynamical masses in many LIRGs/ULIRGs at different redshift ranges (Tacconi et al.~\cite{Tac08}; 
Daddi et al.~\cite{Dad10}; Genzel et al.~\cite{Gen10})

There have also been claims in the literature that the  MW value  of the $M_{\rm dense}$ -- $L'_{\rm HCN(1-0)}$ conversion factor, initially
adopted in Sect.~\ref{SFlaws}, does not apply to mergers either (Gao \& Solomon~\cite{Gao04a}; Graci\'a-Carpio et al.~\cite{Gra08}).  
The high HCN to CO ratios measured in some ULIRGs/LIRGs ($\simeq$0.1) make the requirement that  $\alpha^{\rm CO}_{\rm 
mergers}$$\leq$(1/2--1/4)$\times$$\alpha^{\rm CO}_{\rm MW}$  mostly incompatible with the basic prescription that $M_{\rm dense}$ should not in all likelihood exceed 
0.5$\times$$M_{\rm gas}$ (Gao \& Solomon~\cite{Gao04a}). 

\begin{figure}[th!]
   \centering
   \includegraphics[width=7cm, angle=-90]{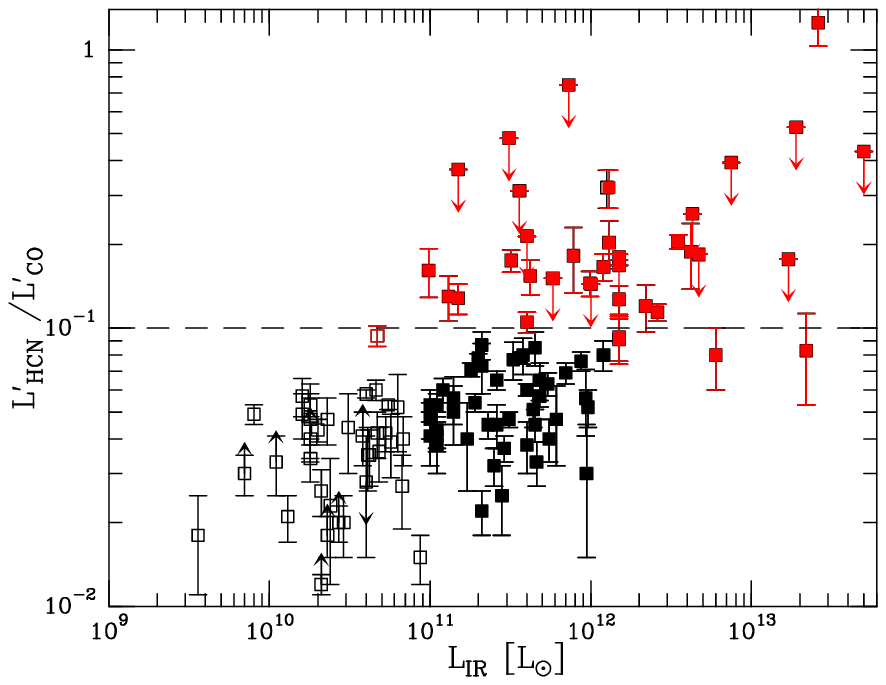}
   \caption{The HCN to CO luminosity ratios as function of $L_{\rm IR}$ for the full sample of galaxies analyzed in this work. Here, normal galaxies  ($L_{\rm IR} < 10^{11} L_{\sun}$) are represented by open squares and IR luminous galaxies  ($L_{\rm IR} > 10^{11} L_{\sun}$) by filled squares. 
   Red markers highlight those galaxies showing over-luminous HCN lines (HCN/CO$\geq$0.1). With the exception of one normal galaxy, all galaxies in this category are IR luminous.}
    \label{lineratios}
\end{figure}

The case of the ULIRG Arp~220 is paradigmatic in this respect. Downes \& Solomon~(\cite{Dow98}) derived from the lower conversion factor for CO a total gas mass for the 
system $M_{\rm gas}$$\sim$5$\times$10$^{9}$M$_{\sun}$. The mass of dense gas derived from our HCN observations if we use the standard 
conversion version factor for HCN ($\alpha^{\rm HCN}_{MW}$) is $M_{\rm dense}$$\sim$1.2$\times$10$^{10}$M$_{\sun}$, i.e., $\sim$(2--3)$\times$$M_{\rm gas}$. This result 
indicates that $\alpha^{\rm HCN}_{Arp220}$ should be $\leq$(1/2-1/3)$\alpha^{\rm HCN}_{MW}$ even in the extreme limit case where $M_{\rm dense}$=$M_{\rm gas}$. 
The new data obtained in our work for other IR luminous galaxies corroborates this picture.  
Figure~\ref{lineratios} shows the HCN to CO ratios as a function of $L_{\rm IR}$ derived for the full sample of 
galaxies used in this paper. This diagram is used to single 
out  {\it over-luminous} HCN line targets in our sample, characterized by HCN/CO$\geq$0.1 within the errors. About 50$\%$ of the 
LIRGs/ULIRGs in our sample are {\it over-luminous} HCN galaxies, identified by the red filled square symbols in Fig.~\ref{lineratios}. By contrast, only one normal galaxy in our sample is {\it over-luminous} in HCN. In these galaxies $\alpha^{\rm HCN}$ has to be lowered by a similar factor as $\alpha^{\rm 
CO}_{\rm mergers}$ relative to the MW canonical values.


Following a similar approach to that employed to constrain the value of  $\alpha^{\rm CO}_{\rm mergers}$, 
Graci\'a-Carpio et al.~(\cite{Gra08}) used the data obtained in a multiline survey of HCN and HCO$^{+}$  of 17 LIRGs and ULIRGs to fit  
$\alpha^{\rm HCN}$ with Large Velocity Gradient (LVG) radiative transfer codes. The main outcome of Graci\'a-Carpio et al.'s study also backs 
up the claim that $\alpha^{\rm HCN}$ is $\sim$3 times lower at high $L_{\rm FIR}$. Lower conversion factor for HCN in extreme starbursts can be related
to the particular chemical environement of the dense molecular gas in these sources where a hot-core like type of chemistry may prevail and enhance HCN abundances (e.g., Lintott et al.~2005).
In addition, non-collisional excitation could also be responsible for the surprising strength of HCN lines in some IR luminous systems (e.g.; Aalto et al.~\cite{Aal95}; Garc{\'{\i}}a-Burillo et al.~\cite{Gar06}; 
Gu\'elin et al.~\cite{Gue07}; Wei{\ss} et al.~\cite{Wei07}). In either case, the HCN conversion factor should be lowered.

The independent lines of evidence mentioned above indicate that for a sizeable fraction of the IR luminous galaxies in our sample, the HCN conversion factor should be lowered. It remains to be proved 
that a similar correction should be applied to all the galaxies in our sample that qualify as IR luminous mergers. With this caveat in mind we re-derive the KS laws in Sect.~\ref{SFlaws-rev} adopting $\alpha^{\rm HCN}$=1/3.2$\times$$\alpha^{\rm HCN}_{\rm MW}$ in LIRGs/ULIRGs. This is 
the same re-scaling factor globally adopted for mergers in the CO study of  Genzel et al.~(\cite{Gen10}).                                                                                                                                    We note that about 70$\%$ of the LIRGs/ULIRGs analyzed in our paper have $L_{\rm IR} > 10^{11.5} L_{\sun}$, the limit beyond which mergers and strongly interacting systems start to dominate over less disturbed disk systems (Sanders \& Ishida~\cite{San04}; Alonso-Herrero et al.~\cite{Alo09}). 
    
\subsection{The L$_{\rm FIR}$--to--SFR conversion factor in normal galaxies} \label{SFR}

  In Sect.~\ref{SFlaws} we used the IR luminosity of the galaxies as a proxy
for their SFR. While this approach is valid for dusty galaxies,  
it has now become clear that in normal star forming galaxies one needs
to account for both the obscured (traced by the IR luminosity) and the
unobscured SFRs (P\'erez-Gonz\'alez et al. ~\cite{Per06}; Calzetti et al.~\cite{Cal07};
Kennicutt et al.~\cite{Ken09}). In particular, combinations of an IR tracer
(e.g., the $24\,\mu$m luminosity)  with the observed H$_{\alpha}$ luminosities
are shown to work well. In this section we evaluate the corrections
needed to account for {\it all} the SFR (obscured+unobscured) 
in our sample of galaxies. 

For the normal star-forming galaxies in our sample we made use of the
SINGS sample data. We have eight galaxies in common with the SINGS
survey. We used the integrated observed (that is, not corrected for
extinction) H$\alpha$ and $24\,\mu$m luminosities
given by Kennicutt et al.~(\cite{Ken09}) to compute the total SFR using the
prescriptions given in their paper. As described in  Kennicutt et al.~(\cite{Ken09}), the
  use of total IR or monochromatic IR luminosities work both equally well to derive
  the total (obscured+unobscured) SFR. On average we find that ${\rm
  SFR}(24\,\mu{\rm m})/{\rm SFR (tot)}=0.5-0.6$.  This indicates that SFR based
  on an IR tracer underestimates the true SFR roughly by a factor of 2 in normal star forming galaxies
  (see also Kennicutt et al.~\cite{Ken09} and references therein).

In LIRGs and ULIRGs, as mentioned above, the IR-based SFR indicators account for 
most of the total SFR. This was shown by recent integral field spectroscopy studies of a sample of more than 50 LIRGs and ULIRGs 
(Garc\'{\i}a-Mar\'{\i}n et al.~\cite{Gar09}; Rodr\'{\i}guez-Zaur\'{\i}n et al.~\cite{Rod11}). Therefore no corrections are applied to derive SFR surface density for LIRGs and ULIRGs.

  We thus adopt in Sect.~\ref{SFlaws-rev} a correction factor of 2 to derive  $\Sigma_{\rm SFR}$ 
from $L^{\rm SFR}_{\rm IR}$ in normal star forming galaxies. This is likely an upper limit to the true correction factor in 
normal galaxies, as we expect that the attenuation will vary more continuously as a function of the FIR luminosity. 
We note however that applying a lower correction factor for normal galaxies  would further reinforce the dual behavior of star formation laws discussed in Sect.~\ref{SFlaws-rev} (see Fig.~\ref{histo-dip}).
 

\begin{figure*}[bth!]
   \centering
         \includegraphics[width=8cm, angle=-90]{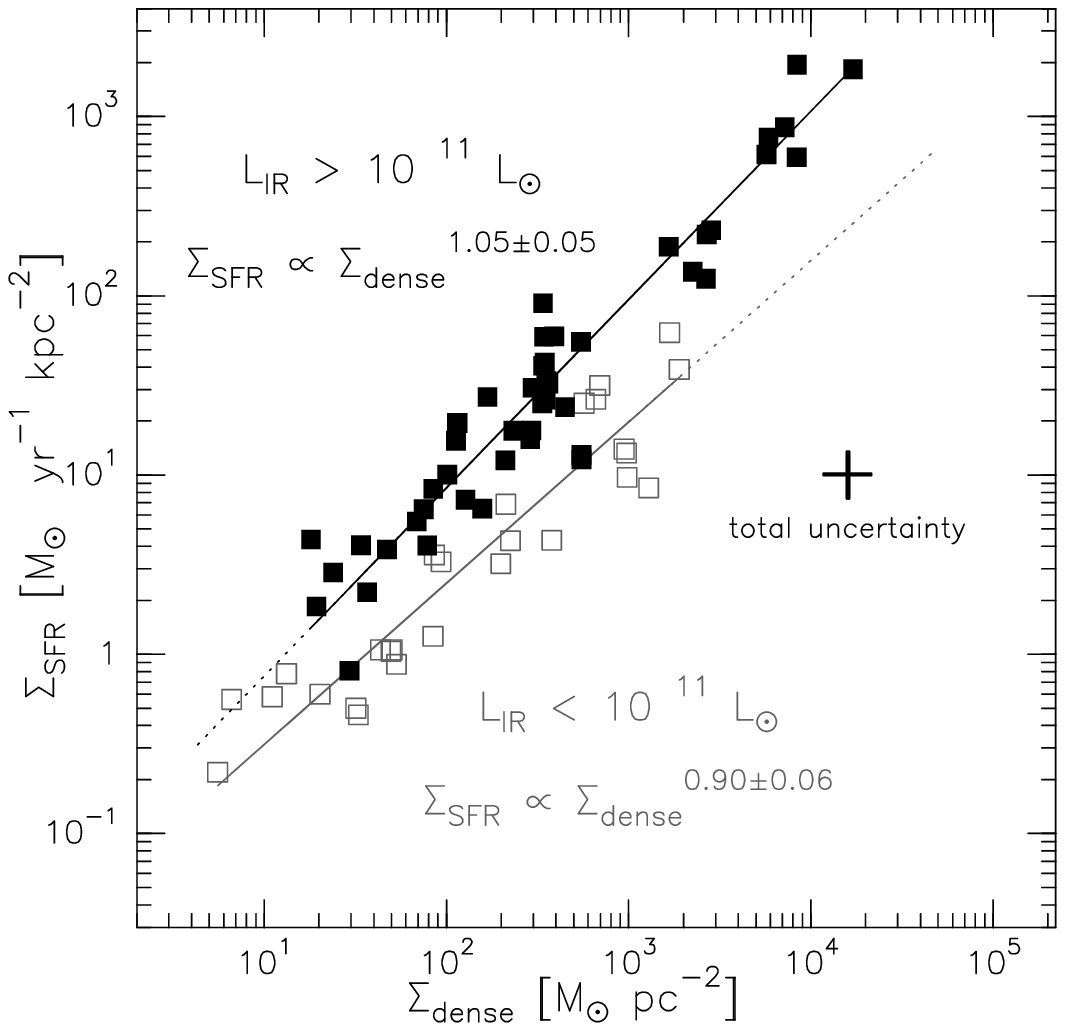}
         \includegraphics[width=8cm, angle=-90]{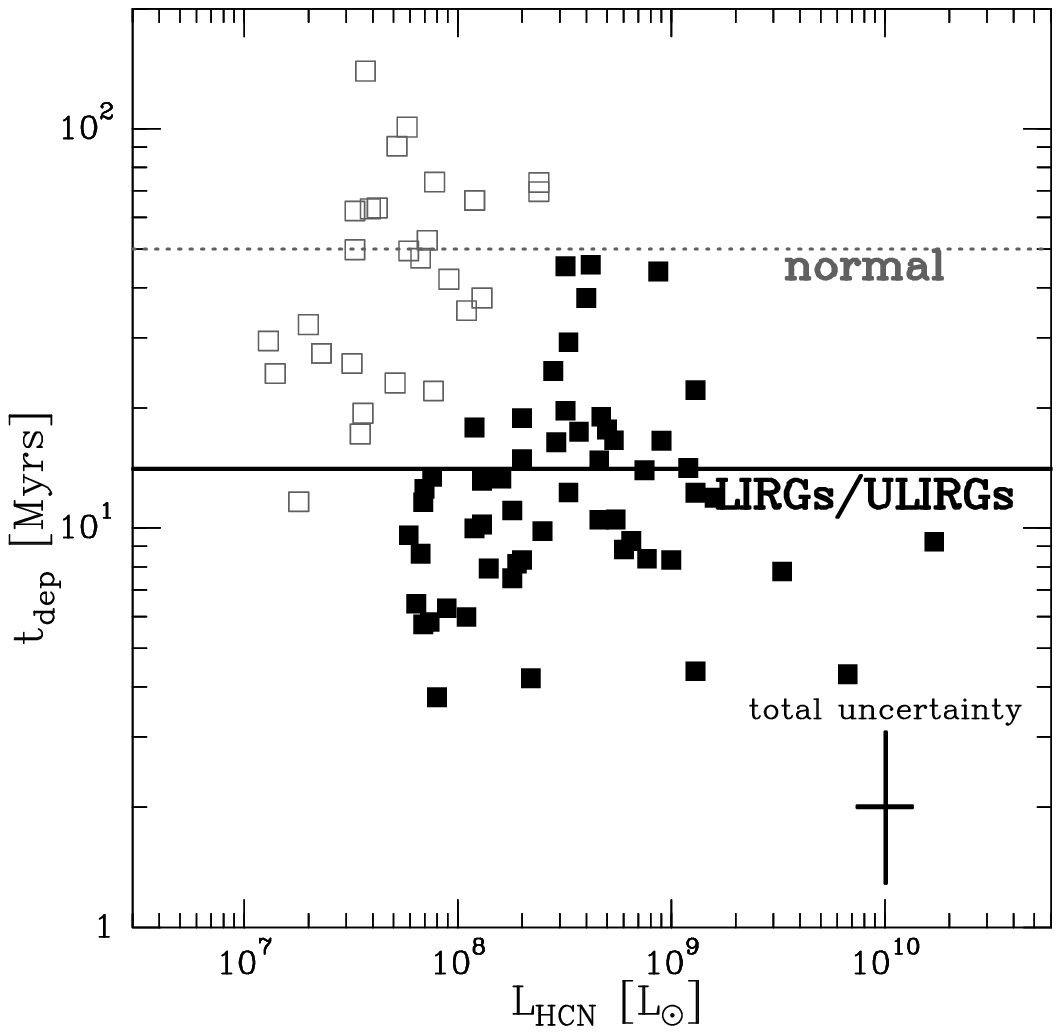}
   \caption{{\bf a)}~({\it left panel}) Same as Fig.~\ref{KS}{\it a)}, but showing the revised two function power law fit  to normal galaxies (grey line) and luminous infrared galaxies (black line) discussed in Sect.~\ref{SFlaws-rev}. Symbols and errorbars as in Fig.~\ref{KS}. {\bf b)}~({\it right panel}) We show the
   revised depletion time scale as a function of $L'_{\rm HCN}$  derived in normal and luminous infrared galaxies, as discussed in Sect. ~\ref{SFlaws-rev}. The dashed and continuous horizontal lines 
 indicate, respectively, the average value of the depletion time scale in normal galaxies (t$_{\rm dep}\sim$50$\pm$5~Myrs) and LIRGs/ULIRGs (t$_{\rm dep}\sim$14$\pm$1.4~Myrs). Symbols and errorbars as in Fig.~\ref{Tdepl}. } 
              \label{KS-revised}
\end{figure*}

 \subsection{Revised star formation laws} \label{SFlaws-rev}

 We have fitted a new two-function power law to the whole sample of galaxies using the revised values of   $\Sigma_{\rm dense}$  (in LIRGs/ULIRGs)
and $\Sigma_{\rm SFR}$ (in normal galaxies)  derived according to Sects.~\ref{ratios} and ~\ref{SFR}. Figure~\ref{KS-revised}{\it a} visualizes the new star formation 
relation in normal galaxies and LIRGs/ULIRGs. The best orthogonal fit solution can be expressed as:

\begin{equation}
  \log{\Sigma_{\rm SFR}} = (0.90 \pm 0.06)\log{\Sigma_{\rm dense}} + (-1.40 \mp 0.15)
\end{equation}

\begin{equation} 
  \mathrm{or\ } \Sigma_{\rm SFR} \simeq 0.04\ \Sigma_{\rm dense}^{0.90}
\end{equation}

\noindent for galaxies with $L_{\rm IR} < 10^{11}\,L_{\sun}$, and:

\begin{equation}
  \log{\Sigma_{\rm SFR}} = (1.05 \pm 0.05)\log{\Sigma_{\rm dense}} + (-1.17\mp 0.16)
\end{equation}

\begin{equation} 
  \mathrm{or\ } \Sigma_{\rm SFR} \simeq 0.07\ \Sigma_{\rm dense}^{1.05}
\end{equation}

\noindent for local and high-$z$ IR luminous galaxies with $L_{\rm IR} \geq 10^{11}\,L_{\sun}$. 

 An inspection of  Fig.~\ref{KS-revised}{\it a} indicates that, compared to Fig.~\ref{KS}~b, the dual behavior is significantly reinforced in the new version of the KS laws.  At the high end of $\Sigma_{\rm dense}$ values ($\sim$a few10$^4$\,M$_{\sun}$pc$^{-2}$)  the normal galaxy law underpredicts $\Sigma_{\rm SFR}$ in IR luminous galaxies by about an order of magnitude, a factor $\sim$8 larger than the typical uncertainty on $\Sigma_{\rm SFR}$. Furthermore, within the range of gas surface densities shared by normal galaxies and LIRGs/ULIRGs, $\Sigma_{\rm dense}$$\sim$2 10$^1$--2 10$^3$\,M$_{\sun}$pc$^{-2}$, the factor 3 to 5 disagreement between the two laws is statistically significant, being a factor $\sim$3--4 larger than the typical uncertainty on $\Sigma_{\rm SFR}$.  Instead, if we try to fit all the observations shown in  Fig.~\ref{KS-revised}{\it a} with a one-parameter law over the whole range of IR luminosities we obtain:

\begin{equation}
  \log{\Sigma_{\rm SFR}} = (1.15 \pm 0.04)\log{\Sigma_{\rm dense}} + (-1.60 \mp 0.06)
  \label{oneparameter}
\end{equation}

This solution has a 2.2 times higher $\chi^{2}$ than the two-function fit found above.

Figure~\ref{KS-revised}{\it b} shows the revised depletion time scales of the dense molecular gas in normal galaxies and LIRGs/ULIRGs. A similar reinforcement of the duality in star formation laws can be identified if we compare the new estimates of t$_{\rm dep}$ with those shown in Fig.~\ref{Tdepl}. The value of t$_{\rm dep}$ is on average a factor 
$\sim$3--4 lower in LIRGs/ULIRGs  (t$_{\rm dep}\sim$14$\pm$1.4~Myrs) compared to normal galaxies (t$_{\rm dep}\sim$50$\pm$5~Myrs). This difference is a factor of 2--3 larger than the total uncertainty of individual data points. 

As extensively discussed in Sect.~\ref{stat-tests}, the revised star formation laws of Eqs.~(15--18), by increasing the spread between normal galaxies and LIRGs/ULIRGs, make more plausible the use of the two-function fit as the best description of the data.


\begin{figure*}[th!]
   \centering
   \includegraphics[width=8cm, angle=-90]{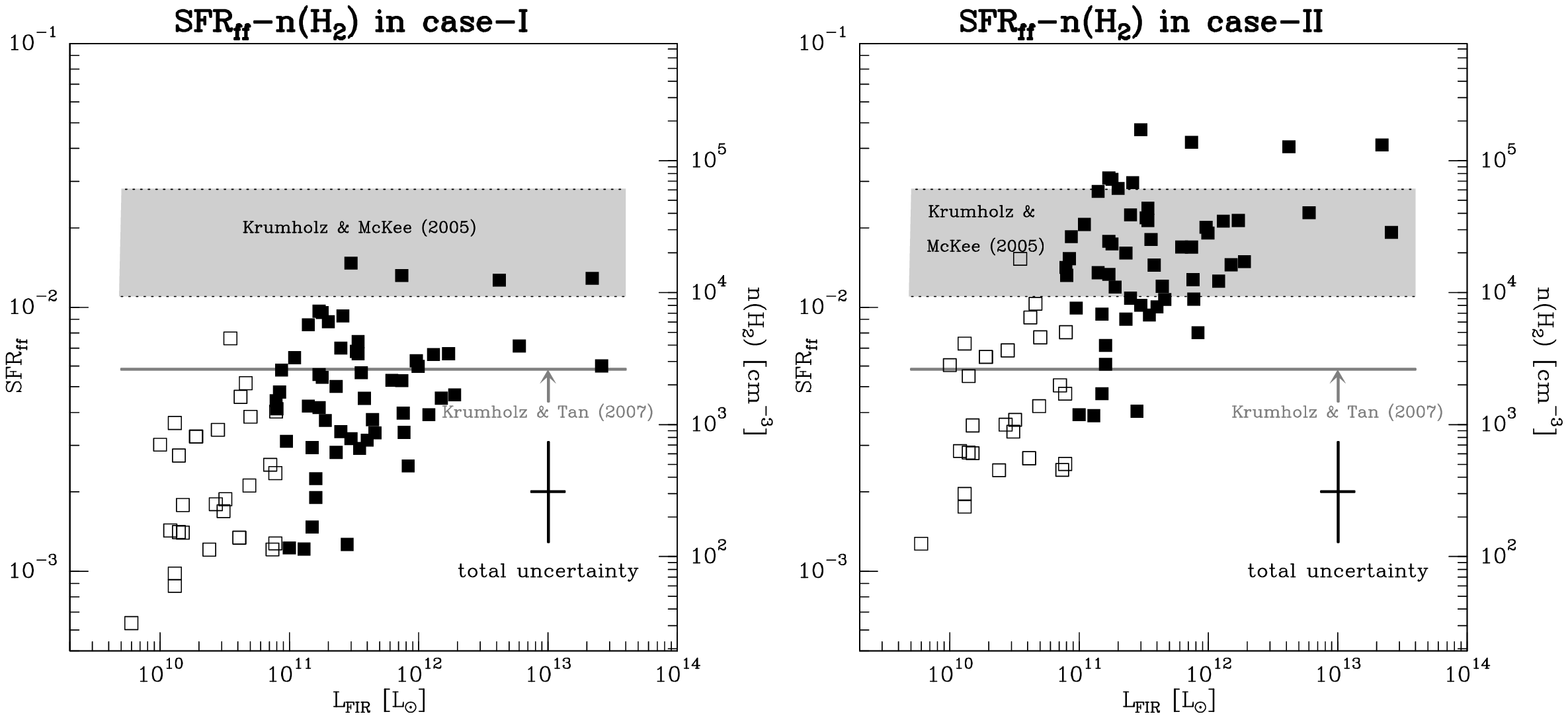}
   \caption{{\bf a)}~({\it left panel}) Star formation rate per free fall time (SFR$_{\rm ff}$; shown in the left Y axis) derived from the observed
    SFE$_{\rm dense}$ in different populations of galaxies assuming a characteristic gas density for the HCN cloud 
    n$_{\rm HCN}$(H$_2$)=3$\times$10$^4$cm$^{-3}$.  We show in the right Y axis the value of n$_{\rm HCN}$(H$_2$)
     derived from SFE$_{\rm dense}$ assuming a constant SFR$_{\rm ff}$=0.02. This value lies within the most likely range for  SFR$_{\rm ff}$$\sim$0.011--0.028 as determined by the star formation model of Krumholz \& McKee~(\cite{Kru05}). 
     We also highlight the average value of SFR$_{\rm ff}$$\sim$0.0058 as determined by  Krumholtz \& Tan~(\cite{Kru07a}) from a compilation of galactic and extragalactic HCN observations. Symbols are as in Fig.~\ref{LFIR-LHCN}.  Errorbars on SFR$_{\rm ff}$ ($\pm$42$\%$) and $L_{\rm FIR}$ ($\pm$30$\%$)  are shown. {\bf b)}~({\it right panel}) Same as {\bf a)} but obtained based on the revised values of $\Sigma_{\rm dense}$  (in LIRGs/ULIRGs) and $\Sigma_{\rm SFR}$ (in normal galaxies)   discussed in Sect.~\ref{factors}. } 
              \label{SFRff}
\end{figure*}

\section{Star formation models: {\it local} vs {\it global}}\label{models}

The evidence of bimodality in star formation laws derived from CO line data is a challenge for star formation models relying only on {\it local} processes (e.g., Genzel et al.~\cite{Gen10}).  Our findings of a similar yet admittedly less extreme duality in the star formation laws derived from HCN, a better tracer of the dense molecular phase, and thus more directly related to star formation, can put stringent constraints on models. By contrast with the highly scattered CO diagram, the smaller intrinsic dispersion in the HCN star formation laws helps assess the verisimilitude of duality with a higher statistical significance.  We discuss in Sect.~\ref{local} to what extent our results based on the HCN observations discussed in this paper can be used to test one of the key predictions of the star formation model of Krumholz \&McKee~(\cite{Kru05}): the constancy of the star formation rate per free-fall time (SFR$_{\rm ff}$). We study in Sect.~\ref{Lratios-models} the values predicted by Krumholz et al.'s models for the $L_{\rm FIR}/L'_{\rm HCN(1-0)}$ luminosity ratios in different galaxy populations and use our data
to benchmark the models. Furthermore, we study in Sect.~\ref{Timescales}  if the inclusion of `global' dynamical time scales offers a better description of the star formation laws derived for the dense gas.

\subsection{Observations vs models: is the star formation rate per free-fall time constant?}\label{local}

Krumholz \&McKee~(\cite{Kru05}) developed an analytic prescription for the star formation laws that are valid for different galactic environments. In their model star 
formation is globally very inefficient as it only takes place in a small subregion of a supersonic turbulent virialized molecular cloud. In this over-dense 
subregion, known as the cloud core(s), gravitational pull is able to take over kinetic energy and the collapse starts. If SFR is the total star formation rate in a galaxy, the star formation rate per free-fall time in objects of class X, SFR$_{\rm ff-X}$, can be obtained as a 
function of SFR and the total mass of X objects, M$_{\rm X}$, (or of their associated variables SFE or t$_{\rm dep}$), as well as of the fraction of star formation occurring in these objects, f$_{\rm X}$ and the 
free-fall time, t$_{\rm ff-X}$:

\begin{equation}
{\rm SFR}_{\rm ff-X} = \frac{f_{\rm X}\ {\rm SFR}\  t_{\rm ff-X}}{M_{\rm X}} = f_{\rm X}\ {\rm SFE}_{\rm X}\ t_{\rm ff-X} = f_{\rm X} \frac{t_{\rm ff-X}}{t_{\rm depl}}
\label{SFRff-X}
\end{equation}

The star formation rate per free-fall time is a function of two parameters: the ratio of the kinetic energy (thermal plus turbulent) to the 
gravitational energy of the cloud ($\alpha_{\rm VIR}$) and the Mach number of the region ($\mathcal{M}$) where $\mathcal{M}^2$ is roughly the ratio of the 
kinetic energy to the thermal energy. One of the key predictions of these models is that the dimensionless  SFR$_{\rm ff}$ is expected to be low and lie within a quite restricted range $\leq$3$\%$ ($\sim$0.011--0.028) for the entire span of plausible values of $\alpha_{\rm VIR}$ ($\sim$1--2) 
and  $\mathcal{M}$ ($\sim$20--80). This offers a quantitative description for why star formation is so inefficient in a highly turbulent molecular medium.

If we take HCN clouds as representative objects then  SFE$_{\rm X}$$\equiv$SFE$_{\rm dense}$$\equiv$1/t$_{\rm dep}$ as defined in Sect.~\ref{SFElaws}. 
We can therefore estimate SFR$_{\rm ff}$  using Eq.~\ref{SFRff-X} directly from observations, provided that we assume a typical density for HCN clouds and 
suppose a value for $f_{\rm X}$. It is reasonable to assume that $f_{\rm X}$ is about unity as all star formation in galaxies is likely taking place at densities 
exceeding the critical  densities of the HCN(1--0) line (n$^{\rm crit}_{\rm HCN}$(H$_2$)$\sim$3$\times$10$^4$cm$^{-3}$/$\tau$, where $\tau$ is the opacity of 
the 1--0 line of HCN; e.g., Gao \& Solomon~\cite{Gao04a}). The actual value of t$_{\rm ff-HCN}$ depends on the density of clouds emitting HCN. 
All in all, observations can be used to verify/falsify the prediction of Krumholz \&McKee~(\cite{Kru05})'s model about the behavior of SFR$_{\rm ff}$.

Krumholz \& Tan~(\cite{Kru07a}) used galactic and extragalactic observations of a variety of objects to derive SFR$_{\rm ff-X}$ as a function of the 
characteristic ISM densities, which  are probed by different molecular tracers (i.e., {\it X-objects}) in their compilation. 
A key outcome of their work is that SFR$_{\rm ff-X}$ seems to stay roughly constant with density. However, the results obtained from HCN suggest
a significant deviation of SFR$_{\rm ff-X}$ from this general trend. The average value of SFR$_{\rm ff}$ derived from the HCN observations used in their paper 
(SFR$_{\rm ff-HCN}\sim$0.0058) lies below the range predicted by Krumholz \& McKee~(\cite{Kru05})'s model. Krumholz \& Tan~(\cite{Kru07a}) 
nevertheless assigned a large error bar to SFR$_{\rm ff-HCN}$,  implicitly assuming that the scattered distribution of SFR$_{\rm ff}$ derived from HCN data is 
mostly random \footnote{Krumholz \& Tan~(\cite{Kru07a}) claim that uncertainties in observations translate into an order of magnitude errorbar.}. 

We use below the HCN data discussed in this work to study the distribution of SFR$_{\rm ff}$ values derived in our sample of galaxies. Our goal is 
to check if the new observations are compatible with the prediction of a constant SFR$_{\rm ff}$. In this comparison  we use the two
versions of the star formation laws derived using the standard conversion factors ({\it Case-I} below) and the revised conversion 
factors discussed in Sect.~\ref{factors} ({\it Case-II} below). As discussed below, exploring different conversion factors is paramount if we are to probe the plausible range of values for SFR$_{\rm ff}$, as these two quantities are interconnected through Eq.~(20).

\subsubsection{Case-I: standard conversion factors}\label{constant}

If we assume that all HCN clouds have densities equal to n$^{\rm crit}_{\rm HCN}$=3$\times$10$^4$cm$^{-3}$ (i.e., we take $\tau$=1) we can obtain 
SFR$_{\rm ff}$ from  Eq.~\ref{SFRff-X}. Figure~\ref{SFRff}{\it a} represents  SFR$_{\rm ff}$ as a function of L$_{\rm FIR}$ obtained from the SFE$_{\rm 
dense}$ values derived in Sect.~\ref{SFlaws}, i.e., prior to the correction of conversion factors discussed in Sect.~\ref{factors}. We note that the distribution of 
SFR$_{\rm ff}$ lies noticeably below the range predicted by the model, as shown in Fig.~\ref{SFRff}{\it a}. In addition to this conspicuous downward shift, the 
deduced SFR$_{\rm ff}$ distribution shows a systematic trend with L$_{\rm FIR}$ and thus cannot  be described as {\it random}. The order of magnitude 
increase in SFR$_{\rm ff}$ from normal galaxies to LIRGs/ULIRGs is statistically significant and it is about a factor of 7 larger than the typical uncertainty of individual data points (42$\%$). The SFR$_{\rm ff}$ progression echoes the SFE$_{\rm dense}$ trend with L$_{\rm FIR}$ 
discussed in Sect.~\ref{SFElaws}. 

To make compatible a constant  SFR$_{\rm ff}$ with the order of magnitude increase in SFE$_{\rm dense}$, the density of HCN clouds has to change notably 
from normal galaxies to LIRGs/ULIRGs, as n$_{\rm HCN}$(H$_2$) scales as $\sim$SFE$_{\rm dense}^2$. We represent  in the right Y axis of Fig.~\ref{SFRff}{\it a} 
the value of n$_{\rm HCN}$(H$_2$) required to fit observations, assuming a fixed SFR$_{\rm ff}$=0.02 for all galaxies, which is close to the average value 
predicted by the star formation model of Krumholz \& McKee~(\cite{Kru05}). The required n$_{\rm HCN}$(H$_2$) densities span two orders of magnitude 
from normal galaxies ($\sim$10$^2$cm$^{-3}$) to LIRGs/ULIRGs ($\sim$10$^4$cm$^{-3}$).  Of note, these densities are exceedingly low 
compared to n$^{\rm crit}_{\rm HCN}$; this poses a problem for the standard scenario of collisional excitation of HCN lines.   Extreme and thus unrealistic high
opacities for HCN would be required to compensate for the low densities: $\tau_{\rm HCN}$=3--300.  As an additional relevant constraint, we would need to 
assume that opacities are a factor 10-100 higher in normal galaxies compared to LIRGs/ULIRGs. This requirement is at odds 
with the commonly measured higher molecular abundances of extreme starbursts (Combes~\cite{Com91}; Wild et al.~\cite{Wil92}; Nguyen et al.~\cite{Ngu92}; 
Krips et al.~\cite{Kri08}; Graci\'a-Carpio et al.~\cite{Gra08}). The current observational 
evidence contradicts the existence of these abnormally low HCN densities, at least in the galaxies for which reliable estimates of n$_{\rm HCN}$(H$_2$) exist 
(e.g., Tacconi et al.~\cite{Tac94}; Sternberg et al.~\cite{Ste94}; Usero et al.~\cite{Use04}; Krips et al.~\cite{Kri08}; Graci\'a-Carpio et al.~\cite{Gra08}).

Based on the new HCN observations presented in this work and the use of standard conversion factors, we derive values for  SFR$_{\rm ff}$ and/or   
n$_{\rm HCN}$(H$_2$) that are well below the theoretical expectations both for normal galaxies and LIRGs/ULIRGs. Furthermore the different densities  required to
fit the two populations of galaxies with a  common constant SFR$_{\rm ff}\sim$0.02 are far from the values derived in observations.   

\subsubsection{Case-II: revised conversion factors}\label{revised}

Figure~\ref{SFRff}{\it b} is similar to  Fig.~\ref{SFRff}{\it a}  but here  SFR$_{\rm ff}$ values have been obtained from the revised estimates of the conversion 
factors discussed in  Sect.~\ref{factors} for $\Sigma_{\rm dense}$  (in LIRGs/ULIRGs) and $\Sigma_{\rm SFR}$ (in normal galaxies). 
Not surprisingly, by adopting the revised conversion factors the global distribution of SFR$_{\rm ff}$ is shifted upward and, at the same time, the
difference between normal galaxies and LIRGs/ULIRGs is increased. Most of the LIRGs/ULIRGs in 
Figure~\ref{SFRff}{\it b} show now values of SFR$_{\rm ff}$ which lie within the range predicted by  Krumholz \& McKee~(\cite{Kru05}) with HCN 
densities very close to  n$^{\rm crit}_{\rm HCN}$. The problem persists however in the fit of normal galaxies, where the predicted values for SFR$_{\rm ff}$ 
and/or  n$_{\rm HCN}$(H$_2$) are still well below the expected range. The paradigm of a common constant SFR$_{\rm ff}$ in all galaxies 
can only be saved if we allow this parameter to be as low as $\sim$0.0035. In this scenario the HCN densities fitting the observed efficiencies would range from n$^{\rm crit}_{\rm HCN}$ in normal galaxies to $\sim$10$^6$cm$^{-3}$ in LIRGs/ULIRGs. 
 The fit in densities nevertheless requires to assume a value for SFR$_{\rm ff}$ that is a factor of 5--6 lower than expected by Krumholz \& McKee' s models  or derived from CO observations (Krumholz \& Tan~\cite{Kru07a}).

We can thus conclude that the paradigm of a constant  SFR$_{\rm ff}$ predicted by models within the range covered by the new HCN observations of normal galaxies and LIRGs/ULIRGs is not supported. The use of revised conversion factors alleviates the problem especially in LIRGs/ULIRGs. However, within the framework of 
{\it local} models it is not possible to fit the observed differences in the SFE$_{\rm dense}$ between normal galaxies and LIRGs/ULIRGs using a common    
constant SFR$_{\rm ff}\sim$0.02 and a set of physically acceptable HCN densities. 


\begin{figure}[th!] 
  \centering 
   \includegraphics[angle=0,width=9cm]{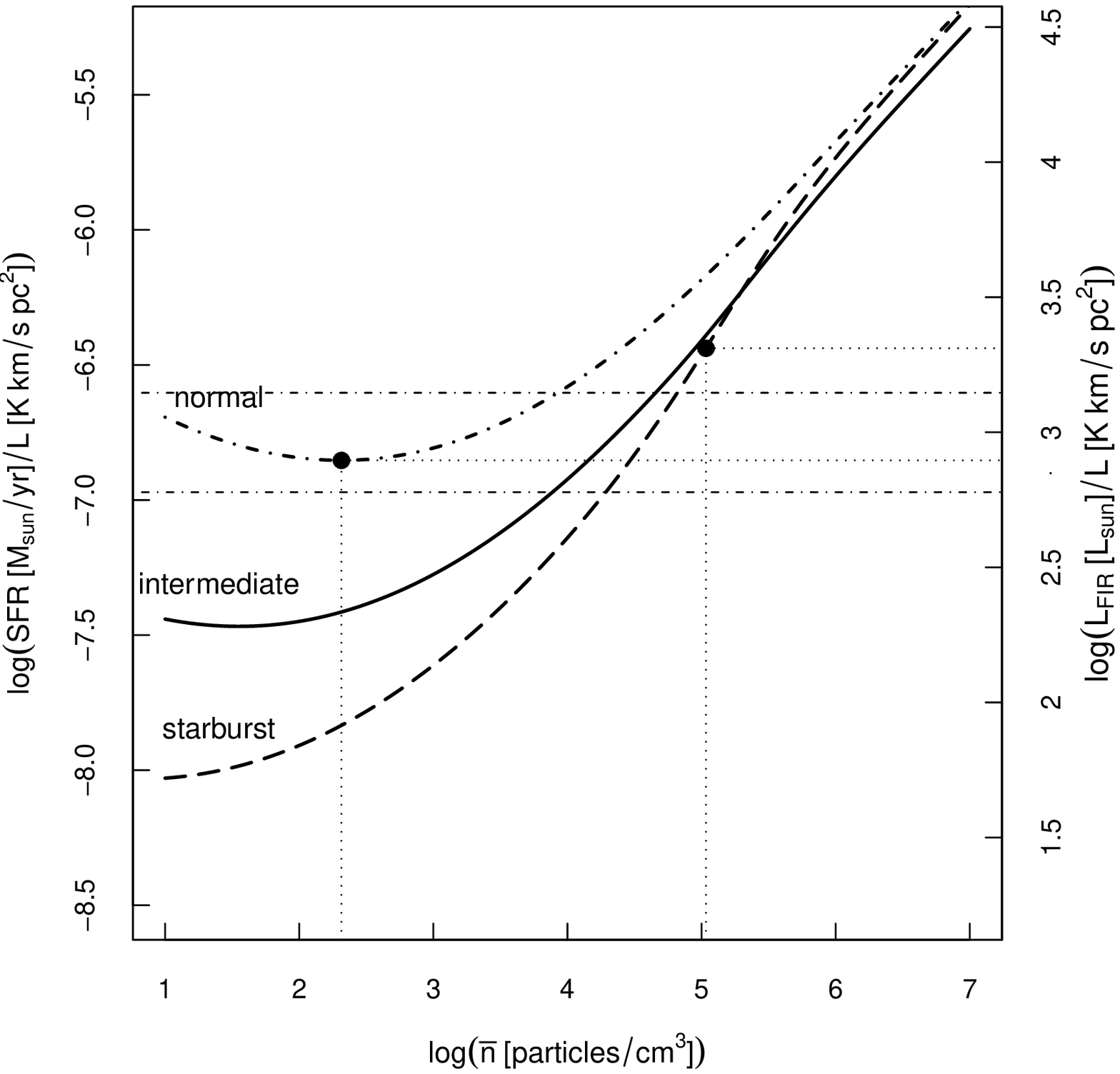}
    \caption{We show the model predicted ratio of the star formation rate or IR luminosity to the HCN(1--0) luminosity as a function of the mean gas density $\bar{n}$ for the three cases examined by Krumholz \& Thompson~(\cite{Kru07b}): normal(dot-dashed line), intermediate(solid line) and starburst(dashed line) galaxies. The dot-dashed horizontal lines highlight the $L_{\rm FIR}/L'_{\rm HCN(1-0)}$ observed
    in normal galaxies and LIRGs/ULIRGs. The vertical dot lines indicate the location of the best agreement between observations and models as discussed in the text.} 
             \label{krumholz} 
\end{figure}



\begin{figure}[th!] 
  \centering 
   \includegraphics[angle=0,width=0.9\hsize]{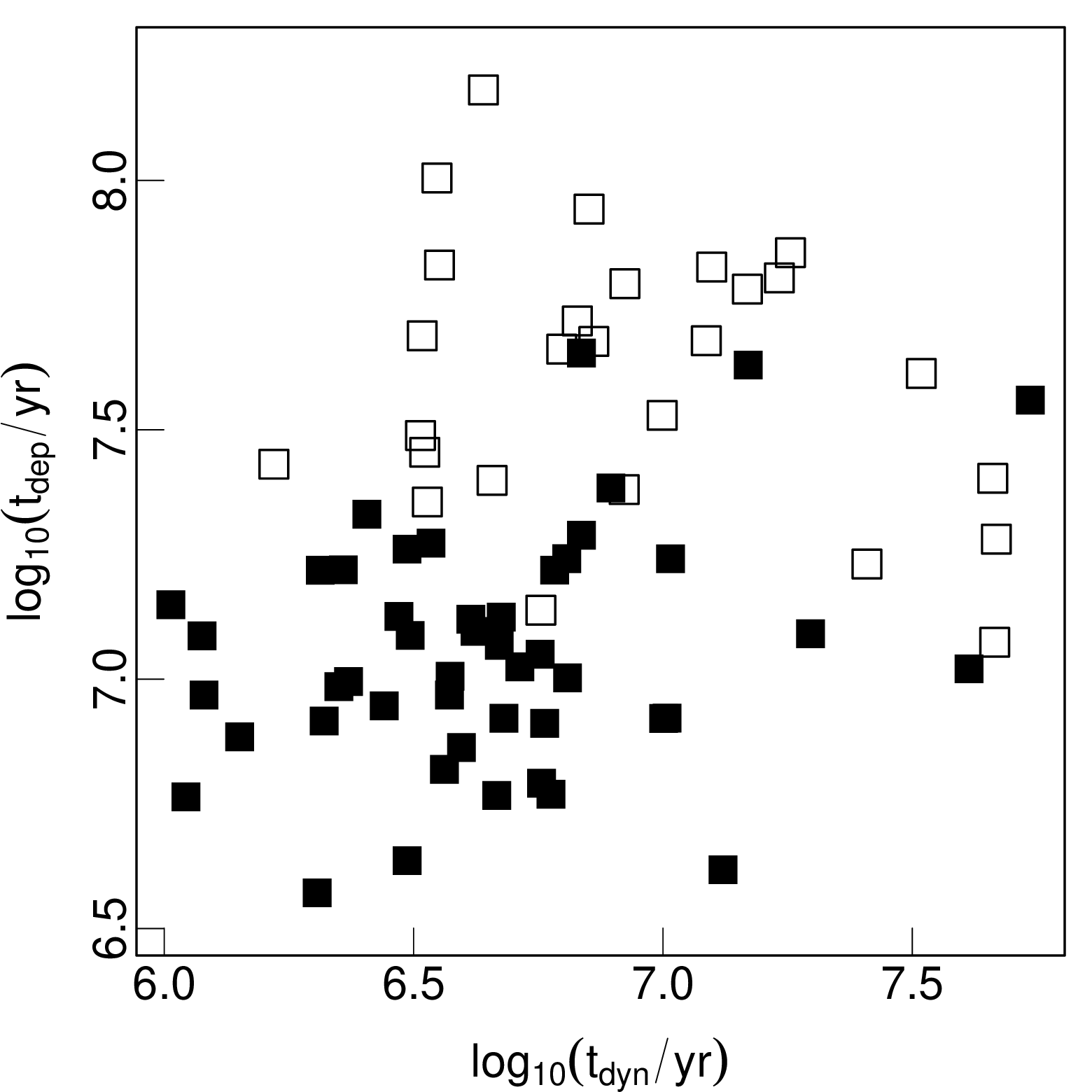}
    \caption{We represent log(t$_{\rm dep}$) versus log(t$_{\rm dyn}$) for normal galaxies ($L_{\rm IR} < 10^{11} L_ 
  {\sun}$, open symbols) and luminous infrared galaxies  ($L_{\rm IR} > 10^{11} L_{\sun}$, filled symbols). } 
             \label{tdep-tdyn} 
\end{figure}


\subsection{Observations vs models: the $L_{\rm FIR}/L'_{\rm HCN(1-0)}$ luminosity ratio}\label{Lratios-models}

Based on the ISM model developed by  Krumholz \& McKee~(\cite{Kru05}), Krumholz \& Thompson~(\cite{Kru07b}) predicted how the $L_{\rm FIR}/L'_{\rm HCN(1-0)}$ luminosity ratios 
should change as a function of the galaxy mean gas density  $\bar{n}$ for three different object classes: normal galaxies, intermediate galaxies and starbursts.
In Krumholz \& Thompson's formulation, molecular clouds in normal galaxies are characterized by low kinetic temperatures (T$_K$=10~K), as well as moderate--to--low Mach numbers ($\mathcal{M}$=30), HCN abundances (X(HCN)=1$\times$10$^{-8}$) and HCN(1--0) opacities ($\tau_{\rm HCN}$=0.5). These molecular cloud parameters are expected to be boosted to higher values in starburst galaxies: T$_K$=50~K, $\mathcal{M}$=80, X(HCN)=4$\times$10$^{-8}$, $\tau_{\rm HCN}$=2 (See Table 1 of Krumholz \& Thompson~\cite{Kru07b} for details). The common envelope to normal and starburst systems is that molecular clouds are  characterized
by a roughly constant dimensionless star formation efficiency measured by SFR$_{\rm ff}$.

One key prediction of this model is that for mean densities $\bar{n}$$\leq$10$^4$cm$^{-3}$, normal galaxies are expected to show higher SFE$_{\rm dense}$ than galaxies classified as starbursts. Another key prediction is that the $L_{\rm FIR}/L'_{\rm HCN(1-0)}$ luminosity ratio should remain roughly constant in a diagram that contains simultaneously data from normal galaxies (characterized by low $\bar{n}$) and starburst galaxies (characterized by high $\bar{n}$) up to densities $\bar{n}$$\leq$10$^4$cm$^{-3}$. A different (less {\it flat}) appearance of SFE$_{\rm dense}$ as a function of $\bar{n}$ can be foreseen at higher densities, however.
Figure~\ref{krumholz}, a modified version of Fig.~2 of Krumholz \& Thompson~(\cite{Kru07b}), purposely explores the predicted SFE$_{\rm dense}$ in the high density regime: 10$^4$cm$^{-3}$$<$$\bar{n}$$<$10$^7$cm$^{-3}$. 
Following the convention of Krumholz \& Thompson~(\cite{Kru07b}) we represent the $L_{\rm FIR}/L'_{\rm HCN(1-0)}$ and SFR/$L'_{\rm HCN(1-0)}$ ratios as a function of the mean gas density  $\bar{n}$ for normal (dot-dashed line), intermediate (solid line) and starburst (dashed line) galaxies\footnote{This figure has been produced using the code of the model made publicly available by Mark Krumholz at http://www.ucolick.org/~krumholz/.}. 
As argued by Krumholz \& Thompson~(\cite{Kru07b}), the models for normal galaxies and starbursts are seen to converge in the high density regime ($\bar{n}$$\geq$10$^5$cm$^{-3}$).

As discussed in Sect.~\ref{SFElaws}, the observations presented in this paper show that the $L_{\rm FIR}/L'_{\rm HCN(1-0)}$ ratio is not constant inside our combined sample: SFE$_{\rm dense}$ is a factor 
$\sim$2--3 higher in LIRGs/ULIRGs  compared to normal galaxies ($<$SFE$_{\rm dense}$$>$(LIRGs/ULIRGs) $\sim$1400$\pm$100$\,L_{\sun}\,{L'}^{-1}$;  $<$SFE$_{\rm dense}$$>$(normal)$\sim$600$\pm$70$\,L_{\sun}\,{L'}^{-1}$). 
These data can be used to benchmark the predictions of the model shown in Figure~\ref{krumholz}. Two conclusions can be drawn from this comparison. First, in relative terms, the model is able to reproduce the increase in
the SFE$_{\rm dense}$ derived from observations as being due to a dramatic increase in the average gas densities. According to Fig.~\ref{krumholz}, if we assume that normal 
galaxies are characterized by $\bar{n}$=2 10$^2$cm$^{-3}$, extreme starbursts (LIRGs/ULIRGs) should be in a $\bar{n}$=10$^5$cm$^{-3}$ regime to fit the observed change in SFE$_{\rm dense}$. 
In absolute terms, the model predicts values of SFE$_{\rm dense}$ for normal galaxies and extreme starbursts that are marginally a factor $\sim$1.5 above the observed values\footnote{We note that the absolute scale along the right Y axis of the Fig.~2 of Krumholz \& Thompson~(\cite{Kru07b}) seems to be incorrect. A re-evaluation of this scale, derived from the output of the model, indicates that the predicted $L_{\rm FIR}/L'_{\rm HCN(1-0)}$ ratios are a factor of 2--3 lower than shown in the mentioned figure. Of particular note, this change helps finding a {\it better} agreement between observations and the model.}. 
This success of the model would be in apparent contradiction with the 
scenario described in the previous section, where we concluded that the star formation efficiency per free-fall time SFR$_{\rm ff}$ in the model, supposed to be $\sim$0.02, has to be lowered by a factor 5--6 to fit the observations (after the revision of conversion factors discussed in Sect.~\ref{revised}). As argued in the following there is no contradiction, however. Based on the output of the model of Krumholz \& Thompson, we can
actually estimate the implicit value of $\alpha^{\rm HCN}$ for the model molecular clouds at densities larger than n$^{\rm crit}_{\rm HCN}$=3$\times$10$^4$cm$^{-3}$. The HCN conversion factor
depends, for each galaxy class, on $\bar{n}$. Assuming the best fit values for $\bar{n}$ reported above, we derive that $\alpha^{\rm HCN}$$\sim$1--2$M_{\sun}\,{L'}^{-1}$ in starbursts and $\sim$3--4$M_{\sun}\,{L'}^{-1}$ in normal 
galaxies in the model, i.e., a factor $\sim$2--3 lower than the values of $\alpha^{\rm HCN}$ adopted in 
Sect.~\ref{ratios}. Taking into account that the best compromise solution for $L_{\rm FIR}/L'_{\rm HCN(1-0)}$ implies ratios that are still $\sim$50$\%$ too high (Fig.~\ref{krumholz}), we conclude that the model (with SFR$_{\rm ff}$$\sim$0.02 and our preferred values for $\alpha^{\rm HCN}$) is a factor of $\sim$4 off. This is similar to the factor of $\sim$5--6 disagreement found in Sect.~\ref{revised}.

The conclusion common to the present section and Sect~\ref{local} is that a rough agreement between the HCN observations presented in this paper and the predictions of {\it local} models can only be reached if SFR$_{\rm ff}$$\sim$0.005-0.01 and/or if $\alpha^{\rm HCN}$ is still a factor of $\sim$a few lower than our favored values for normal galaxies and starbursts (LIRGs/ULIRGs). In qualitative agreement with this picture, the recent paper of Krumholz et al.~(\cite{Kru11}) has presented conclusive evidence that the best correspondence between the prediction of {\it local} models and the results of CO galaxy surveys requires lowering SFR$_{\rm ff}$ to 0.01, i.e., a factor of 1.5--3  below the value previously assumed by  Krumholz \& McKee~(\cite{Kru05}).


\begin{figure}[th!] 
  \centering 
   \includegraphics[angle=-90,width=0.9\hsize]{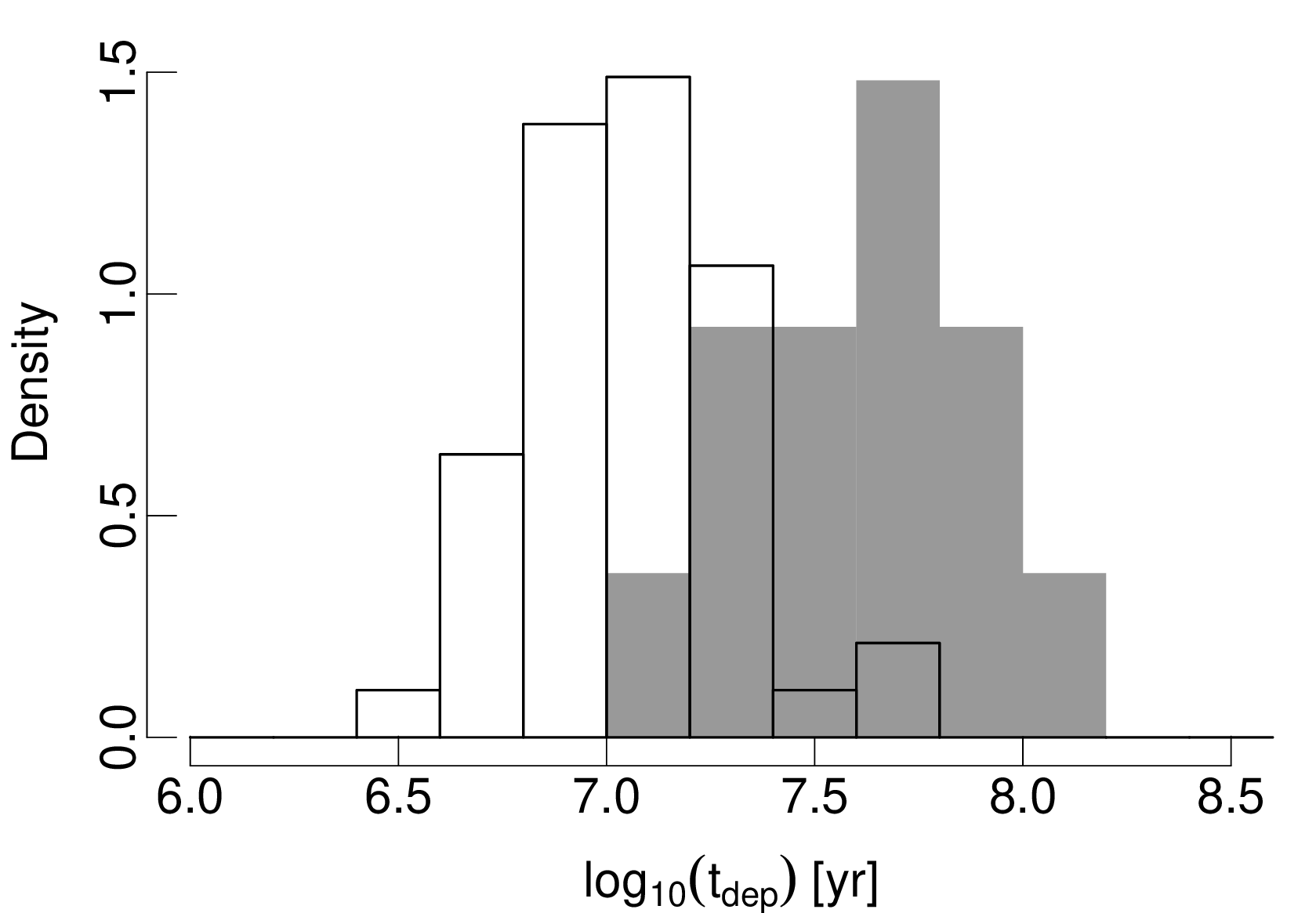}\\ 
   \includegraphics[angle=-90,width=0.9\hsize]{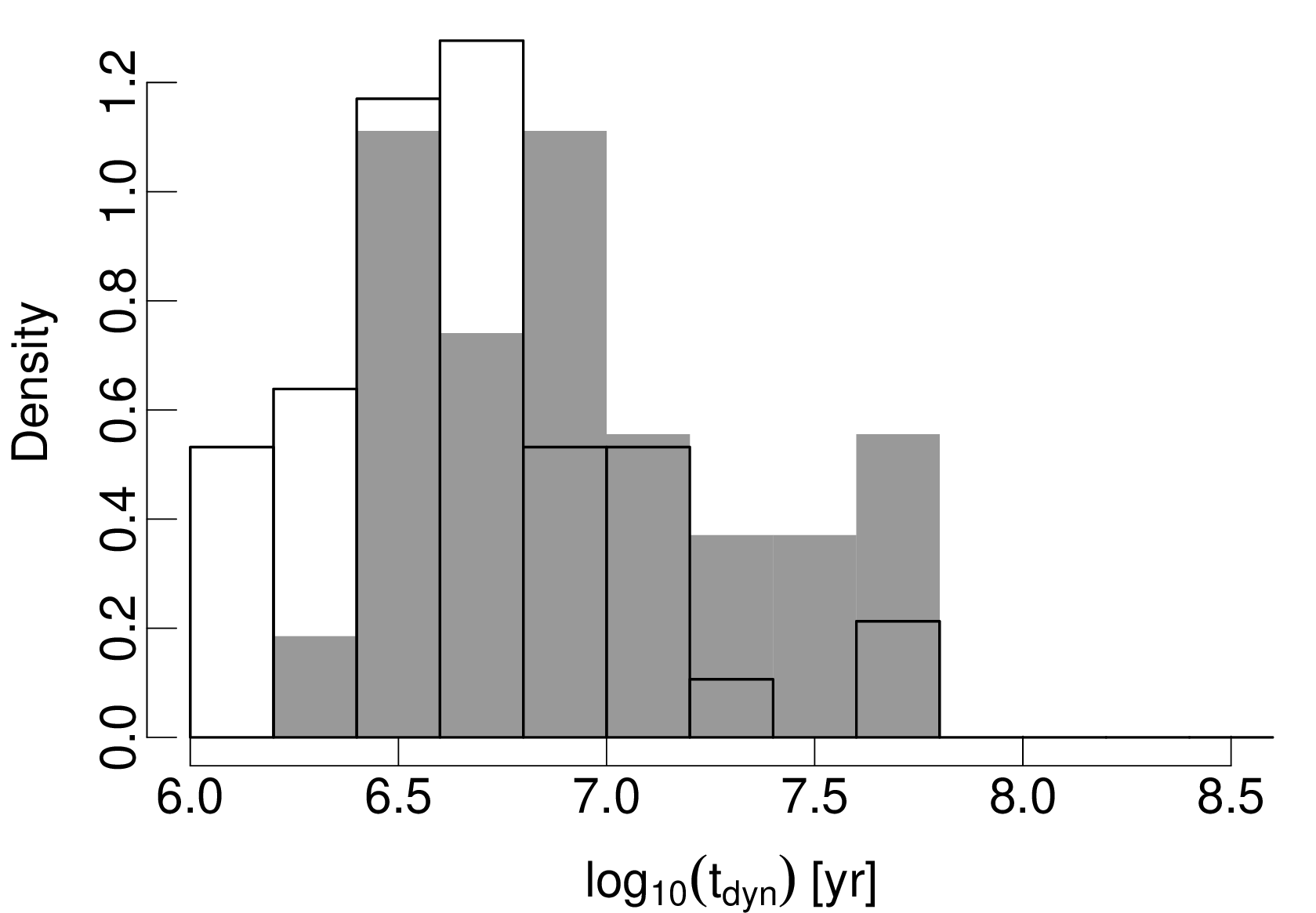}\\ 
   \includegraphics[angle=-90,width=0.9\hsize]{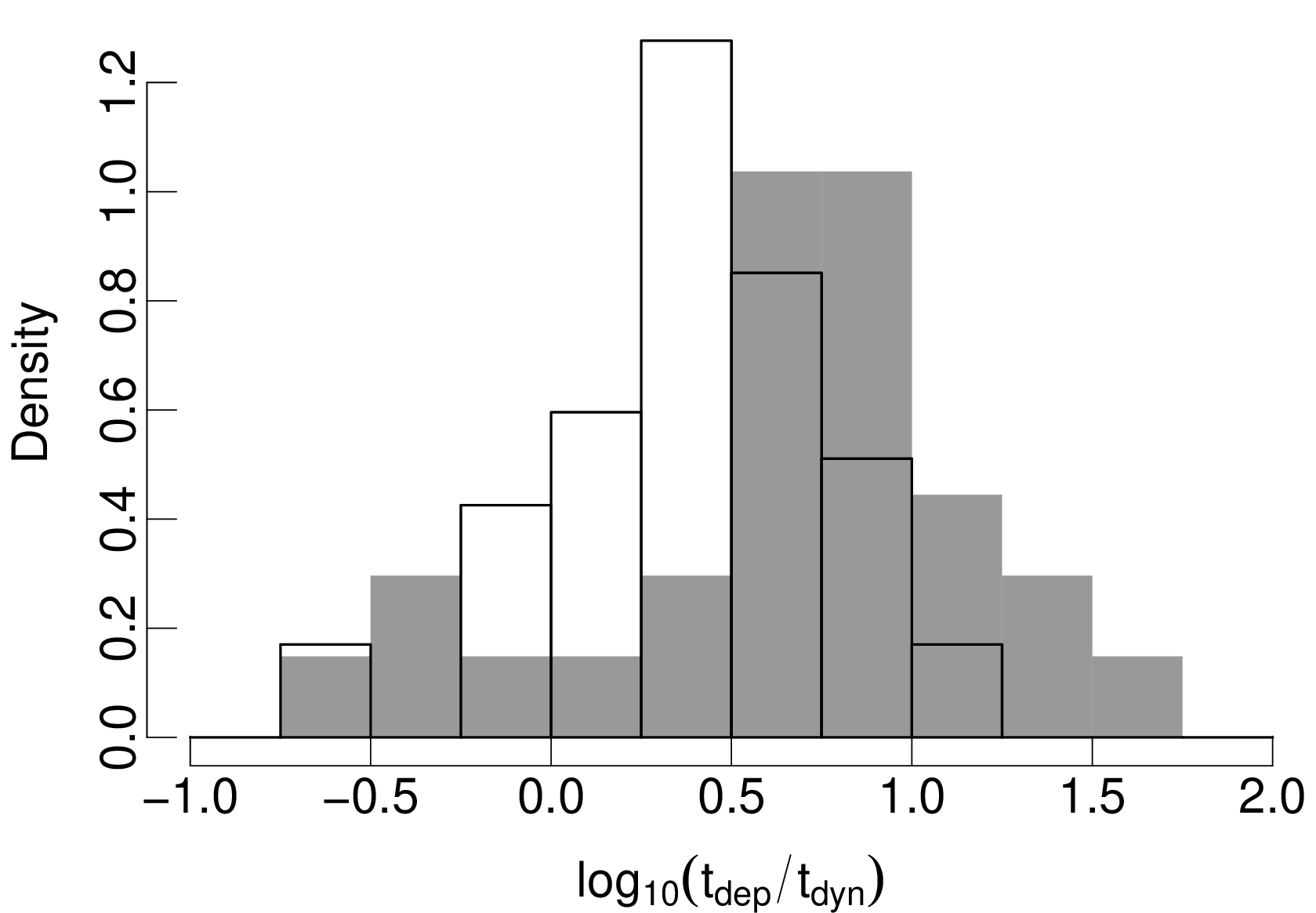} 
  \caption{{\bf a)}~({\it upper panel}) Histograms showing the number density of galaxies (number of galaxies per unit interval, as shown) as a function of log(t$_{\rm dep}$) for normal galaxies ($L_{\rm IR} < 10^{11} L_ 
  {\sun}$, grey filled histogram) and luminous infrared galaxies  ($L_{\rm IR} > 10^{11} L_{\sun}$, empty histogram). {\bf b)}~({\it middle panel}) Same as  {\bf a)} but 
  representing the number density of galaxies as a function of log(t$_{\rm dyn}$).  {\bf c)}~({\it lower panel}) Same as  {\bf a)} but representing the 
  number density of galaxies as a function of log(t$_{\rm dep}$/t$_{\rm dyn}$). } 
             \label{times} 
\end{figure}



\begin{figure}[bth!]
   \centering
   \includegraphics[width=9.5cm, angle=0]{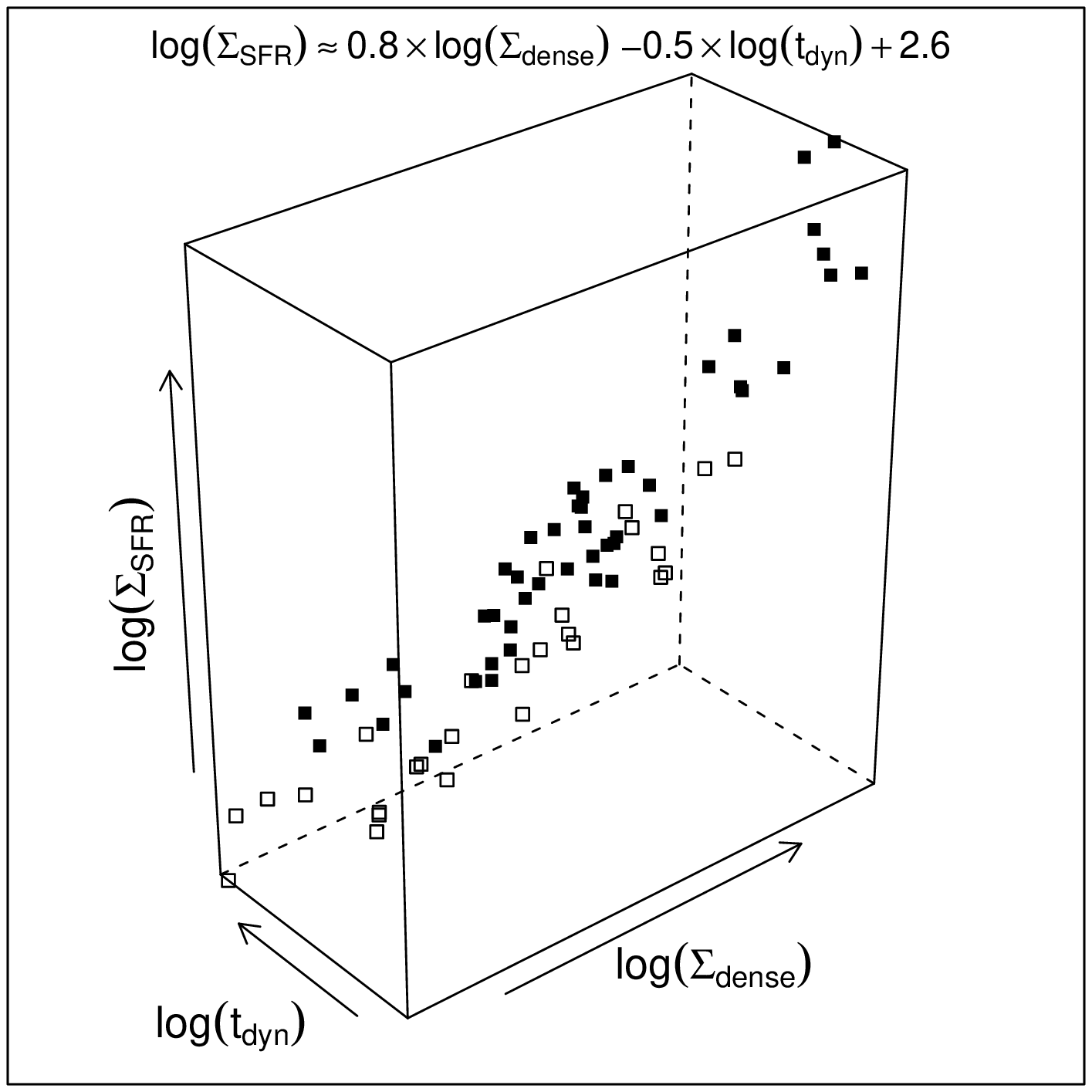}
   \caption{Three-dimensional representation of the bi-parametric star formation law giving the star formation rate density $\Sigma_{\rm SFR}$ as a function
   of the dense molecular gas surface density $\Sigma_{\rm dense}$  and the dynamical time t$_{\rm dyn}$.} 
              \label{KS3d}
\end{figure}

\subsection{Observations vs models: is `global' dynamical time scale a key parameter?}\label{Timescales}

 We have provided evidence that the conversion of dense molecular gas into stars is more efficient in LIRGs/ULIRGs than in normal galaxies. 
Although this has been ascribed to the existence of very different local physical properties (densities) of the clouds in these two populations of galaxies, 
we have seen in Sects~\ref{local} and \ref{Lratios-models} that this hypothesis has difficulties in explaining the HCN results discussed in this paper (unless
significantly lower SFR$_{\rm ff}$ and/or $\alpha^{\rm HCN}$ are assumed).  Alternatively, the star formation rates could be 
regulated by the global large-scale dynamics of the ISM,  more perturbed in strongly interacting galaxies (in our sample most of them are LIRGs/ULIRGs) than in normal galaxies. In this 
scenario we expect that the differences in SFE$_{\rm dense}$ (or equivalently in t$_{\rm dep}$) between the two populations of galaxies can be reproduced by  
similar differences in  t$_{\rm dyn}$, as  SFE$_{\rm dense}$$\equiv$1/t$_{\rm dep}$$\propto$1/t$_{\rm dyn}$. To explore this possibility, we compare in this section the distribution of dynamical time scales and depletion time scales of the dense molecular gas in our sample.

Figure~\ref{tdep-tdyn} represents  log(t$_{\rm dep}$) versus log(t$_{\rm dyn}$) for normal galaxies and LIRGs/ULIRGs. Depletion time scales of the 
dense gas t$_{\rm dep}$ are derived as defined in Sect.~\ref{SFElaws}. Dynamical time scales t$_{\rm dyn}$ are defined as a crossing time scale unit
determined by the ratio between the radius of the star forming region, where the HCN emitting gas is expected to lie, and half the full width of the HCN line. 
When available, we have corrected t$_{\rm dyn}$  for the value of the inclination of the galaxies.  Whereas there is a clear segregation between the two populations of galaxies as to the values of t$_{\rm dep}$, the corresponding differences in t$_{\rm dyn}$ are less significant.

Figure~\ref{times} shows a histogram representation of the number density of galaxies as a function of t$_{\rm dep}$, t$_{\rm dyn}$ and the t$_{\rm dep}$/t$_{\rm dyn}$ ratio. 
We split our sample into normal galaxies  ($L_{\rm IR} < 10^{11} L_ {\sun}$, grey filled histograms) and luminous infrared galaxies  
($L_{\rm IR} > 10^{11} L_{\sun}$, empty histograms). Depletion times (top panel) roughly span an order of magnitude in each subsample. They peak 
around $\sim$10~Myr in the case of LIRGs/ULIRGs  and around $\sim$50~Myr in the case of normal galaxies. The clear separation between the two 
histograms makes apparent the higher star formation efficiency of LIRGs/ULIRGs. The middle panel shows the dynamical time scales of the studied galaxies. The two histograms 
extend across a common range of $\sim$1.5~orders of magnitude, although the dynamical time scales in LIRGs/ULIRGs are about a factor of 2 lower on 
average. The existence of differences  between the two populations of galaxies can be expected, for many LIRGs/ULIRGs are mergers where the interaction has 
brought most of the molecular gas to smaller radii. We note however that the differences reported for t$_{\rm dyn}$ fall short of explaining the factor 5 difference 
in t$_{\rm dep}$.

We represent in the lower panel of Fig.~\ref{times} the histograms of the t$_{\rm dep}$/t$_{\rm dyn}$ ratio. This is equivalent to the ratio between the 
mass of dense molecular gas and the mass of stars  formed in one dynamical time scale. Although the shape (intrinsic dispersion) of the two histograms is 
similar, t$_{\rm dep}$/t$_{\rm dyn}$ ratios in LIRGs/ULIRGs are typically a factor of 2 lower than in normal galaxies. Taken at face value, this suggests that the 
amount of dense gas converted into stars per unit of dynamical time is somewhat higher in LIRGs/ULIRGs. This result would contradict the idea that star 
formation is governed mainly by global dynamical time scales, as in this case we expect  that the t$_{\rm dep}$/t$_{\rm dyn}$ ratio is roughly constant, thus 
showing no significant differences between the two galaxy populations. 

The inability of `global' star formation models to fully explain the KS law derived from HCN, anticipated above, is better illustrated in Fig.~\ref{KS3d}. We show in this 
figure the three-dimensional representation of the bi-parametric star formation law that expresses the star formation rate density $\Sigma_{\rm SFR}$ as a 
function of the dense molecular gas surface density $\Sigma_{\rm dense}$  and the dynamical time t$_{\rm dyn}$.  The best fit to the HCN data is given by:

\begin{equation}
  \log{\Sigma_{\rm SFR}} = (0.8\pm0.1)\log{\Sigma_{\rm dense}} + (-0.5\pm0.2)\log{t_{\rm dyn}} + (2.6 \mp 1)
   \label{twoparameter}
\end{equation}

While the power law index for $\Sigma_{\rm dense}$  (0.8$\pm$0.1) is not far from unity within the errors, the corresponding index for 
t$_{\rm dyn}$ (--0.5$\pm$0.2) is not compatible with --1. This is against the theoretical expected power law index for 
$\Sigma_{\rm dense}$/t$_{\rm dyn}$  ratio that should be about $\sim$1 (Kennicutt~\cite{Ken98}). Furthermore, the goodness of the fit of Eq.~\ref{twoparameter}, estimated by a standard  $\chi^{2}$ analysis, indicates a meager 12$\%$ improvement with respect to the value of  $\chi^{2}$ obtained in the single parameter fit of Eq.~\ref{oneparameter}. With the 
inclusion of `global' dynamical time scales we are not able to significantly improve the fit of star formation laws derived from HCN.


Several biases in the way the dynamical time scales are derived could explain why the t$_{\rm dep}$/t$_{\rm dyn}$ histograms shown in Fig.~\ref{times} are not coincident for the two 
populations of galaxies analyzed in this paper. In particular, dynamical time scales in normal galaxies, where star formation is typically much more widely spread in the disks compared to LIRG/ULIRGs, could be underestimated.  However LIRGs/ULIRGs do not behave as a monolithic family in this respect: some LIRGs, especially in the range L$_{\rm IR}\leq$10$^{11.4}$L$_{\odot}$ can have spread-out star formation (Hattori et al.~\cite{Hat04}; Alonso-Herrero et al.~\cite{Alo06}; Rodr\'{\i}guez-Zahur\'{\i}n et al.~\cite{Rod11}). 

%
%
%
%
%

%

\section{Summary and conclusions}\label{summary}

We have used the IRAM 30m telescope to observe a sample of 19 LIRGs in the 1--0 lines of CO, HCN and HCO$^+$. The galaxies have been extracted from a sample of local LIRGs with available high-quality and high-resolution images obtained at optical, near and mid IR wavelengths, which probe the star formation activity. The new data presented 
in this work allow us to expand the number of LIRGs studied in HCN-HCO$^+$ lines by more than a factor 3 when compared to previous works. The chosen LIRG sample has a range of HCN 
luminosities that overlaps to a significant extent with that of normal galaxies. This is a key to testing if LIRGs represent a transition point in star formation (SF) laws between normal galaxies 
and ULIRGs. In particular we study if the {\it observed} bimodality of star formation laws in galaxies derived from CO line data in previous surveys, can be extended to the  higher density regime probed by HCN. We use these results to explore the validity of different theoretical 
prescriptions for star formation laws where local and/or global dynamical time scales are explicitly included.

We summarize below the main results and conclusions of this work:

\begin{itemize}

\item

We find that the star formation efficiency of the dense gas (SFE$_{\rm dense}$) estimated by  $L_{\rm FIR}/L'_{\rm HCN(1-0)}$ 
luminosity ratios is on average a factor $\sim$2--3 higher in LIRGs/ULIRGs ($\sim$1400$\pm$100$\,L_{\sun}\,{L'}^{-1}$)  compared to normal galaxies ($\sim$600$\pm$70$\,L_{\sun}\,{L'}^{-1}$).  
With the addition of the new LIRG sample we confirm on a more solid statistical basis that the  $L_{\rm FIR}$-$L'_{\rm HCN(1-0)}$ correlation is significantly 
super-linear (power law $\sim$1.23$\pm$0.05) when we include normal galaxies and LIRGs/ULIRGs in the fit.

\item

 We have derived KS laws relating the two surface densities, $\Sigma_{\rm SFR}$ and $\Sigma_{\rm dense}$ in normal galaxies and LIRGs/ULIRGs.  A fit to the full  sample gives a power index $N = 1.12 \pm 0.04$. However, LIRGs/ULIRGs and normal galaxies are not fully overlapping in this scatter plot.  In order 
 to quantify if a dual fit applies to the power law, we have split the sample into normal and IR luminous galaxies. While the ($\Sigma_{\rm SFR}$, $\Sigma_{\rm dense}$) distribution cannot be described as strictly bimodal, we find that a two-function power law fit  with indexes close to unity qualifies as a much better description of the star formation relation for the dense gas. This result is seen to be solid against the statistical biases inherent to this analysis.

\item

If we account for the different conversion factors for 
 HCN in extreme starbursts and for  the unobscured star formation rate in normal galaxies, for which we find new evidence in these observations, the duality in star formation laws is enhanced. This result extends 
 to the higher molecular densities probed by HCN lines the more extreme bimodal behavior of star 
 formation laws, derived from CO molecular lines by two recent surveys (Daddi et al.~\cite{Dad10}; Genzel et al.~\cite{Gen10}).  The revised depletion time scales for 
 the dense molecular gas show a  significant difference  between LIRGs/ULIRGs  (t$_{\rm dep}\sim$14$\pm$1.4~Myrs) and normal galaxies (t$_{\rm dep}\sim$50$\pm$5~Myrs).

\item   
 We have confronted the results  of the new HCN observations with the predictions of theoretical models in which the efficiency of star formation is determined by the ratio 
 of a \emph{constant} star formation rate per free-fall time (SFR$_{\rm ff}$) to the local free-fall time (t$_{\rm ff}$)  (Krumholz \&McKee~\cite{Kru05}).  Within the framework of these models we find that it is possible to fit the observed differences in the SFE$_{\rm dense}$ between normal galaxies and LIRGs/ULIRGs using a common constant SFR$_{\rm ff}$ and a set of physically acceptable HCN densities, but only if SFR$_{\rm ff}$$\sim$0.005--0.01 and/or if $\alpha^{\rm HCN}$ is still a factor of $\sim$a few lower than our favored values.

\item

We have studied if the inclusion of `global' dynamical time scales offers a better description of the star formation relations derived for the dense gas.  This approach helps finding a marginally better universal relation for the dense molecular gas. In the best fit 
solution for $\Sigma_{\rm SFR}$$\propto$$\Sigma_{\rm dense}^{n}$ t$_{\rm dyn}^{m}$, the power law index for $\Sigma_{\rm dense}$  ($n$=0.8$\pm$0.1) is not far from unity within the errors, i.e., in agreement with the theoretical expected value.
Nevertheless, the corresponding index for t$_{\rm dyn}$ ($m$=--0.5$\pm$0.2) is not compatible with --1, a result whose theoretical implications remain to be understood. 
   
\end{itemize}

A fit of the SF laws similar to the one discussed  in Sect.~\ref{Timescales} was first explored by Genzel et al.~(\cite{Gen10}), who used CO line data in a sample of normal SF galaxies and extreme starbursts at different redshift ranges. The two-parameter (M$_{\rm gas}$, t$_{\rm dyn}$) SFR law derived by Genzel et al.~(\cite{Gen10}) (SFR$\propto$M$_{\rm gas}^{1.4}$ t$_{\rm dyn}^{-0.8}$) is successful at providing a universal SF relation. Of particular note, the power law indexes in their fitted SF relation are {\it also} found to deviate from the theoretical expected values (SFR$\propto$M$_{\rm gas}^{1}$ t$_{\rm dyn}^{-1}$), however. Altogether the SF relation obtained  in Eq.~\ref{twoparameter} from HCN data and the one obtained by  Genzel et al.~(\cite{Gen10}) from CO are very similar to the KS laws obtained, respectively, by Gao \& Solomon~(\cite{Gao04b}) and Kennicutt~(\cite{Ken98}) {\it provided that we remain in the 
$\Sigma_{\rm SFR}$-$\Sigma_{\rm (dense)gas}$ projection of the SF laws}. 

The bottom line, however, is that the different dynamical time scale ranges found for normal galaxies and extreme starbursts are not sufficient to dissolve the duality observed in the $\Sigma_{\rm SFR}$-$\Sigma_{\rm (dense)gas}$ projection of the SF laws (see also Krumholz et al.~\cite{Kru11}). 
A possible explanation for this failure can be searched for in the uncertainties that plague the definition and the observational determination of t$_{\rm dyn}$. This is paramount if we are to compare systems that are known to be in very different dynamical states. So far the most widely accepted definition for t$_{\rm dyn}$, also used in this work, is equivalent to that of a crossing or orbit time scale for the system, which is determined from observations. However, while normal galaxies can be described as dynamically relaxed rotating disks, mergers are strongly interacting and dynamically very disturbed. In this context using a common definition for t$_{\rm dyn}$ is questionable. High-spatial resolution observations can help disentangle the role of large-scale dynamics in the small-scale spatially resolved SF laws both in normal galaxies and in extreme starbursts.

Alternatively, the inclusion of global dynamical time scales might not be the key to estimating the correct scaling of the SF relation and thus determining the absolute value of the SFR for different galaxy populations.
From the observational point of view there is ample room for improvement in our understanding of how conversion factors for different tracers of molecular gas may differ in galaxies. As discussed above, this is a prerequisite to accurately calibrating SF relations and exploring their differences between galaxies. So far, the evidence for different conversion factors in normal galaxies and extreme starbursts is limited to CO and HCN lines. Multi-transition studies to be done for different molecular tracers will be paramount to tackling this issue.

 Besides providing a better handle of conversion factors, required to convert intensities (I) into gas surface densities ($\Sigma_{\rm gas}$), multi-line observations can yield an estimate of volume gas densities ($\rho_{\rm gas}$). This will ease a more direct comparison between observations and models that predict how the SFR volume density should depend on $\rho_{\rm gas}$, rather than on $\Sigma_{\rm gas}$. In this context it is worthy of note that Krumholz et al.~(\cite{Kru11}) have recently argued that a simple, local, volumetric star formation law with SFR$_{\rm ff}\sim$0.01 can dissolve the apparent bimodality found by Genzel et al.~(\cite{Gen10}) if different scale-heights  are assumed for the gas disks (h$_{\rm gas}$) in normal galaxies and extreme starbursts. While this is a plausible explanation, we note that the observational constraints on how h$_{\rm gas}$ might differ in different galaxy populations are currently scarce. The change in h$_{\rm gas}$ required to make duality/bimodality dissapear in SF laws remains thus to be observationally validated.

\begin{acknowledgements}
	 We acknowledge the IRAM staff at 30m telescope for their help during the observations 
	 data reduction.  SGB acknowledges support from MICIN within program CONSOLIDER INGENIO 2010, under grant 'Molecular Astrophysics: The 
	 Herschel and ALMA Era--ASTROMOL' (ref CSD2009-00038). AAH, MPS, LC and SA acknowledge support from  AYA2010-21161-C02-1. MPS also acknowledges support from the CSIC under grant JAE-Predoc-2007. We also thank the anonymous referee for his/her insightful and constructive report.
      
\end{acknowledgements}

\end{document}